\documentclass[twocolumn]{aastex62}
\pdfoutput=1 
\usepackage{amsmath,amstext}
\usepackage[T1]{fontenc}
\usepackage{apjfonts}
\usepackage{comment}
\usepackage{footnote}
\usepackage{booktabs}
\usepackage[switch]{lineno} 
\usepackage[referable]{threeparttablex}
\shorttitle{10 years of GBM pulsars}
\shortauthors{The GAPP team}

\begin{document}
\title{The Ups and Downs of Accreting X-ray Pulsars: \\
Decade-long observations with the $\emph{Fermi}$ Gamma-Ray Burst Monitor}

\author[0000-0002-0380-0041]{C.~Malacaria}
\affiliation{NASA Marshall Space Flight Center, NSSTC, 320 Sparkman Drive, Huntsville, AL 35805, USA}\thanks{NASA Postdoctoral Fellow}
\affiliation{Universities Space Research Association, Science and Technology Institute, 320 Sparkman Drive, Huntsville, AL 35805, USA}
\author{P.~Jenke}
\affiliation{University of Alabama in Huntsville, NSSTC, 320 Sparkman Drive, Huntsville, AL 35805, USA}
\author{O.J.~Roberts}
\affiliation{Universities Space Research Association, Science and Technology Institute, 320 Sparkman Drive, Huntsville, AL 35805, USA}
\author[0000-0002-8585-0084]{C.A.~Wilson-Hodge}
\affiliation{NASA Marshall Space Flight Center, Huntsville, AL 35812, USA}
\author{W.H.~Cleveland}
\affiliation{Universities Space Research Association, Science and Technology Institute, 320 Sparkman Drive, Huntsville, AL 35805, USA}
\author{B.~Mailyan}
\affiliation{University of Alabama in Huntsville, NSSTC, 320 Sparkman Drive, Huntsville, AL 35805, USA}
\collaboration{on behalf of the GBM Accreting Pulsars Program Team}
\noaffiliation


\begin{abstract}
We review more than 10 yr of continuous monitoring of accreting X-ray pulsars with the all-sky Gamma-ray Burst Monitor (GBM) aboard the Fermi Gamma-Ray Space Telescope.
Our work includes data from the start of GBM operations in 2008 August, through to 2019 November. Pulsations from 39 accreting pulsars are observed over an energy range of $10-50\,$keV by GBM. The GBM Accreting Pulsars Program (GAPP) performs data reduction and analysis for each accreting pulsar and makes histories of the pulse frequency and pulsed flux publicly available. We examine in detail the spin histories, outbursts, and torque behaviors of the persistent and transient X-ray pulsars observed by GBM. The spin period evolution of each source is analyzed in the context of disk-accretion and quasi-spherical settling accretion-driven torque models. Long-term pulse frequency histories are also analyzed over the GBM mission lifetime and compared to those available from the previous Burst and Transient Source Experiment (BATSE) all-sky monitoring mission, revealing previously unnoticed episodes in some of the analyzed sources (such as a torque reversal in 2S 1845--024). We obtain new, or update known, orbital solutions for three sources. Our results demonstrate the capabilities of GBM as an excellent instrument for monitoring accreting X-ray pulsars and its important scientific contribution to this field.
\end{abstract}

\section{Introduction}

Accreting X-ray Pulsars (XRPs) were discovered almost $50\,$yr ago, when X-ray pulsation was detected from Cen~X-3 and Her~X-1 \citep{Giacconi1971, Tananbaum1972}, and subsequently interpreted as a rotating, magnetized Neutron Star (NS) accreting the stellar wind expelled by a donor companion star \citep{PringleRees1972,DavidsonOstriker1973,LPP1973}.
For magnetized NSs, where the magnetic field strength is $B\sim10^{12}\,$G, the stellar wind flow is disrupted by the magnetic pressure and channeled to the magnetic polar caps, the so-called ``hotspots''.
Here, the potential energy is converted into X-ray radiation with a luminosity L$_{acc}$ of:
\begin{equation}
L_{acc} \approx \frac{G\,M_{NS}\,\dot{m}}{R_{NS}}
\end{equation}
where  $M_{\rm NS}=1.4\,M_{\sun}$ and $R_{\rm NS}=10\,$km are the mass and radius, respectively, of a typical NS, and $\dot{m}$ is the mass accretion rate.

The study of an XRP system consisting of a magnetized, accreting compact object and an optical companion is key to understanding the behavior of matter under extreme conditions and for probing the evolutionary paths of both the binary system and its individual components. The NS represents the final stage of the evolutionary track of massive stars as a supernova remnant, characterized by extreme densities, high magnetic and gravitational fields, and a very small moment of inertia. Therefore, knowledge is required from multiple scientific disciplines in order to describe NSs in an astrophysical context (i.e., accretion processes, plasma physics,  nuclear physics,  electrodynamics, general relativity, and quantum theory). Furthermore, the presence of a donor companion makes these systems excellent laboratories for the study of additional astrophysical processes, such as the stellar wind environment, radiative effects, and matter transfer. Finally, the presence of an orbiting XRP makes these objects invaluable tools for the characterization of orbital elements and component masses.


The Milky Way and the Magellanic Clouds contain $\sim230$ XRPs\footnote{\url{http://www.iasfbo.inaf.it/~mauro/pulsar_list.html}}.
Recent reviews on XRPs and their observational properties are given in \citet{Caballero+Wilms12}, \citet{Walter+Ferrigno16}, \citet{Maitra17}, \citet{Paul+17}.
These systems have pulse periods that range from several ms to hours. For approximately half of those systems, their X-ray activity has only been observed serendipitously during transient episodes. 

In this paper, we review more than 10 yr of observations of XRPs with the Gamma-Ray Burst Monitor (GBM), an all-sky, transient monitor aboard the Fermi observatory. The wide field of view and high timing capability of GBM are particularly suited to the continuous study of both transient and persistent XRPs. As of 2019 November, GBM has detected a total of $39$ XRPs. 

This paper is organized as follows.
In Sect.~\ref{sec:physics} we briefly review the accretion physics onto magnetized compact objects.
In Sect.~\ref{sec:gbm} we describe the GBM instrument and its data handling.
In Sect.~\ref{sec:timing} we describe the timing analysis applied to the GBM raw data in order to obtain its data products.
In Sect.~\ref{sec:overview} we give an overview of the binary systems observed by GBM and the type of X-ray activity that characterizes those systems.
In Sect.~\ref{sec:individuals} we describe each XRP system, providing a summary of their spin history as seen previously by other observatories and currently by GBM.
Finally, in Sect.~\ref{sec:discussion} we discuss the main results from the population study and from single systems in the most interesting cases.
We summarize the importance of the GBM Pulsar Project in Sect.~\ref{sec:summary}.

As the estimation of the spectral luminosity and measurement of the distance from the sources are important aspects of this work, we describe them separately in the Appendices~\ref{sec:bolo} and ~\ref{sec:gaiadist}, respectively.

\section{Accretion Physics onto Magnetized Neutron Stars}\label{sec:physics}

When accretion occurs onto a magnetized NS, the accreted material does not flow smoothly onto the surface of the compact object but is mediated by the NS's magnetic field \citep{PringleRees1972}.
At a certain distance from the NS surface, namely at the Alfv\'{e}n radius $r_{\rm A}$, the energy density of the magnetic field balances the kinetic energy density of the infalling material:

\begin{equation}
r_A= \left(\frac{\mu^4}{2GM\dot{M}^2}\right)^{1/7}=6.8\times10^{8}\,\dot{M}_{10}^{-2/7}\,\mu^{4/7}_{30}\,M_{1.4}^{-1/7}\,cm
\end{equation}
where $\dot{M}_{10}$ is the accreted mass in units of $10^{-10}\,M_\odot\,$yr$^{-1}$, $\mu_{30}$ is the magnetic moment in units of $10^{30}\,$G\,cm$^{3}$ (corresponding to a typical magnetic field strength of $10^{12}\,$G at the NS surface), and $M_{1.4}$ is the mass of the NS in units of $1.4\,M_\odot$.
The material can then penetrate the NS magnetic field via Rayleigh-Taylor and Kelvin-Helmholtz instabilities and when accretion is mediated through an accretion disk via magnetic field reconnection with small-scale fields in the disk and turbulent diffusion (\citealt{Arons76, GLa,Kulkarni08}, and references therein).
In a disk, the magnetic threading produces a broad transition zone composed of two regions: a broad outer zone, where the disk angular velocity is nearly Keplerian, and a narrow inner zone or boundary layer, where the disk angular velocity significantly departs from the Keplerian value.
The outer radius of the boundary layer is identified as the magnetospheric radius $r_{\rm m}$:

\begin{equation}
r_m = k\,r_A
\end{equation}

where the dimensionless parameter $k$, also called the coupling factor, is ${\sim}0.5$ as given by \citet{GLa}, but it ranges from $0.3$ to $1$ in later models (\citealt{Wang96, Li97, Li+Wang99, Long+05, Bessolaz08, Zanni13, Dallosso16}, and references therein). It can be more generally considered as a function of the accretion rate, $k(\dot{M})$, and of the inclination angle between the neutron star rotation and magnetic field axes, and it can be significantly smaller than that obtained in the model by \citet[see, e.g., \citealt{Bozzo09}]{GLa}.

Disk-driven accretion can be inhibited by a centrifugal barrier if the pulsar magnetosphere rotates faster than the Keplerian velocity of the matter in the disk. This condition is realized when the inner disk radius, coincident with the magnetospheric radius $r_{m}$, is greater than the co-rotation radius, $r_{c\rm o}$, at which the Keplerian angular disk velocity $\omega\r_{\rm co}$ is equal to the angular velocity of the NS,  $\sqrt[]{GM_{NS}/r_{co}}$:

\begin{equation}
r_{co}=\left(\frac{GM_{NS}\,P_{s}^2}{4\pi^2}\right)^{1/3}
\end{equation}

where $\omega=2\pi/P_s$ is the rotational frequency of the NS, and $P_{\rm s}$ is the NS spin.
For a standard NS mass of $1.4\,M_\odot$, the co-rotation radius is of the order of $r_{\rm co}=1.7\times10^8\,P_{\rm s}^{2/3}\,$cm.


The relative positions of these radii determine the accretion regime at work, driven by the possible onset of (either a magnetic or centrifugal) barrier that inhibits direct wind accretion, also called the ``gating'' mechanism \citep{Illarionov75, Stella85, Bozzo08}.
Different regimes of plasma cooling also play a role in quasi-spherical wind-accretion onto slowly rotating NSs \citep{Shakura12, Shakura13,Shakura17}. In these systems, a hot shell forms above the NS magnetosphere and, depending on the mass accretion rate, can enter the magnetosphere either through inefficient radiative plasma cooling or by efficient Compton cooling. At the same time, the plasma mediates the angular momentum removal from the rotating magnetosphere by large-scale convective motions.

When $r_{\rm m}>r_{\rm co}$, the centrifugal barrier rises and mass is propelled away or halted at the boundary, rather than being accreted. This carries angular momentum from the NS, which consequently begins to spin down, and conditions become favorable for accretion via the propeller mechanism as the star enters this regime~\citep{Illarionov75}. However, for $r_{\rm m}<r_{\rm co}$, matter and thus angular momentum, is transferred to the spinning NS. Accordingly, in the case of disk accretion, the total torque $N$ that the disk exerts on the NS is composed of two terms:

\begin{equation}\label{eq:torque1}
N = N_0 + N_{mag}.
\end{equation}

where $N_0=\dot{M}\,\sqrt[]{(G\,M\,r_{\rm m})}$ is the torque produced by the matter leaving the disk at r$_{\rm m}$ to accrete onto the NS, while $N_{mag}=-\int_{r_{\rm m}}^{\infty}B_\phi\,B_z\,r^2\,dr$ is the torque generated by the twisted magnetic field lines threading the disk outside r$_{\rm m}$.
Following \citet{GLa,GLb}, the total torque in Eq.~\ref{eq:torque1} can also be expressed as

\begin{equation}
N = n(\omega_s)\,\dot{M}\,\sqrt[]{G\,M\,r_{co}},
\end{equation}

where $n$ is a dimensionless torque that is a function of the fastness parameter $\omega_s$:
\begin{equation}
\omega_s=\frac{\nu_s}{\nu_k}=\left(\frac{r_{in}}{r_{co}}\right)^{3/2},
\end{equation}

where $\nu_{s}$ and $\nu_k$ are the spin frequency and the Keplerian frequency, respectively.
Accretion from a disk leads to a spin period derivative (\citealt{GL77,GLa,GLb}) equal to
\begin{equation}
-\dot{P}=5.0\times10^{-5}\,\mu^{2/7}_{30}\,n(\omega_s)\,R_{NS_6}^{6/7}\,M_{1.4}^{-3/7}\,I_{45}^{-1}\,P_s^2\,L^{6/7}_{37}\, s\,s^{-1}.
\end{equation}

where $R_{\rm NS_6}$ is the NS radius in units of $10^6\,$cm, I$_{45}$ is the NS moment of inertia in units of $10^{45}\,$g\,cm$^{2}$, and L$_{37}$ is the bolometric luminosity in the X-ray band (1-200 keV) in units of $10^{37}\,$erg\,s$^{-1}$.

For $0<\omega_s<0.9$, a good approximation of the dimensionless torque is \citep{Klus13b}:

\begin{equation}
n(\omega_s) = 1.4\,(1-2.86\,\omega_s)\,(1-\omega_s)^{-1},
\end{equation}

which, for $M_{\rm{NS}}=1.4\,M_\odot$ and $R_{\rm{NS}}=10\,$km, results in a torque \citep{Ho14} of 

\begin{equation}\label{eq:torque2}
-\dot{P}=7.1\times10^{-5}\,\textrm{s\,yr}^{-1}\,k^{1/2}\times(1-\omega_s)\,\mu^{2/7}_{30}\,(P\,L^{3/7}_{37})^2.
\end{equation}

Eq.~\ref{eq:torque2} is widely used in the literature to study accretion disk related phenomena of spin derivatives observed in accretion XRPs.

On the other hand, the quasi-spherical accretion model has been introduced to explain the behavior of wind-accreting systems that show long-term spin period evolution \citep{Gonzalez+12,Gonzalez-Galan18, Shakura12, Postnov2015GX304}. 
This model describes two different accretion regimes, separated by a critical mass accretion rate $y=\dot{M}/\dot{M}_{cr}$, corresponding to a luminosity of $4\times10^{36}\,$erg\,s$^{-1}$.
At lower luminosities, an extended quasi-static shell is formed by the matter that is gravitationally captured by the NS and that is subsonically settled down onto the magnetosphere. 
The quasi-static shell mediates the exchange of angular momentum between the captured matter and the NS magnetosphere by turbulent viscosity and convective motions. 
Both spin-up and spin-down are possible in the subsonic regime, even if the specific angular momentum of the accreted matter is prograde.
As the accretion rate increases above the critical value, the flow near the Alfv\'{e}n surface becomes supersonic and a freefall gap appears above the magnetosphere due to the strong Compton cooling, causing the accretion to become highly unstable.
In this regime, depending on the sign of the specific angular momentum, either spin-up or spin-down is possible.

The quasi-spherical accretion model also takes into account the coupling of the rotating matter with the magnetosphere at different regimes. 
A strong coupling regime is realized for rapidly rotating magnetospheres, in which the exchange of angular momentum between the accreted matter and the NS can be described as
\begin{equation}
    I\dot{\omega} = K_{mag} + K_{surf}
\end{equation}{}

where $I$ is the NS's moment of inertia, $K_{\rm mag}$ is the contribution to the spin frequency evolution brought by the plasma-magnetosphere interactions at the Alfv\'{e}n radius and can be either positive or negative, and $K_{\rm surf}$ is the spin-up contribution due to the angular momentum returned by the matter accreted onto the NS (see Eq.s~17-19 in \citealt{Shakura12}).
In the moderate coupling regime, a similar relation holds with different coupling coefficients (see Eq.s~27-29 in \citealt{Shakura12}). To determine the main dimensionless parameters, the model was used to fit observations from a few long-period pulsars.
Accordingly, the spin-down rate $\dot{\omega}_{\rm sd}$ (where $\omega = 2\pi\nu$ is the angular frequency) observed in those systems is \citep{Postnov2015GX304}
\begin{equation}\label{eq:QSAMdown}
\begin{split}
\dot{\omega}_{sd} \approx 10^{-8} [Hz\, d^{-1}]\, \Pi_{sd}\,\mu_{30}^{13/11}\,\left(\frac{\dot{M}}{10^{16}\,g/s}\right)^{3/11}\,\left(\frac{P_s}{100 s}\right)^{-1}\, 
\end{split}
\end{equation}

where $\Pi_{\rm sd}$ is a parameter of the model, usually in the range $5--10$, $\dot{M}$ is the mass accretion rate normalized for a typical luminosity of $10^{37}\,$erg\,s$^{-1}$ (assuming $L=0.1\dot{M}c^2$), $P_{\rm s}$ is the pulsar spin period, and $\mu_{30}$ is the magnetic moment $BR^3$ in units of $10^{30}$ G\,cm$^3$. The spin-up rate $\dot{\omega}_{\rm su}$ is
\begin{equation}\label{eq:QSAMup}
\begin{split}
\dot{\omega}_{su} \approx 10^{-9} [Hz\, d^{-1}]\, \Pi_{su}\,\mu_{30}^{1/11}\,\left(\frac{P_{orb}}{10\,d}\right)^{-1}\, \left(\frac{\dot{M}}{10^{16}\,g/s}\right)^{7/11}
\end{split}
\end{equation}
where $P_{\rm orb}$ is the binary orbital period, and $\Pi_{\rm su}\approx\Pi_{\rm sd}$ \citep{Shakura14a}.

\section{The Fermi GBM X-Ray Monitor}\label{sec:gbm}

GBM is an unfocused, background-dominated, all-sky instrument aboard the Fermi Gamma-ray Space Telescope \citep{Meegan2009}.
It consists of $14$ uncollimated, inorganic scintillator detectors: $12$ thallium-doped sodium iodide (NaI) detectors and two bismuth germanate (BGO) detectors. The NaI detectors have an effective energy range of $\approx8\,$keV$-1\,$MeV, while the BGOs cover an energy range from $\approx200\,$keV$-40\,$MeV. As the emission of accreting pulsars is dominant only below $\sim100\,$keV, data from the BGO detectors will not be used in this work. The NaI detectors are arranged into four clusters of three detectors, placed around each corner of the spacecraft in such a fashion that any source un-occulted by the Earth will illuminate at least one cluster. 

GBM has three continuous (C) data types: CTIME data, with a nominal time resolution of $0.256\,$s and eight energy channels used for event detection and localization, CSPEC data, with a nominal time resolution of $4.096\,$s and $128$ energy channels used for spectral modeling, and continuous time tagged event (CTTE) data with timestamps for individual photon events ($2\,\mu$s precision) over $128$ energy channels. The latter has been available since 2012 November. 

Even though GBM is devoted to hunting Gamma-ray Bursts, it has proven to be an excellent tool in the monitoring of other transient X-ray sources as well. Consequently, the GBM Accreting Pulsars Program\footnote{\url{https://gammaray.nsstc.nasa.gov/gbm/science/pulsars.html\#}} (GAPP) was developed, with the aim of analyzing pulsars detected by GBM. In the context of the GAPP, two different pulse search strategies have been implemented: the daily blind search and the targeted (i.e. source-specific) search. The blind search consists of computing the daily fluxes for $26$ directions ($24$ equally spaced on the Galactic plane, plus the Magellanic Clouds), using CTIME data type. For each direction, a blind Fast-Fourier Transform (FFT) search is performed between $1\,$mHz and $2\,$Hz (and up to 40 Hz with CTTE data). This ensures sensitivity to new sources, new outbursts from known sources, and pulsars whose pulse period is poorly constrained. Typically, only the first three GBM CTIME channels are used for this search: channels 0 ($8-12\,$keV), 1 ($12-25\,$keV), and 2 ($25-50\,$keV). When a new source is detected through the blind search, its galactic longitude is interpolated from several directions, with the strongest signals in the power spectrum obtained by the FFT technique. 
The targeted search consists of an epoch-folding-based search over much smaller frequency ranges than the blind search method, which sometimes includes a search over the frequency derivative (see also Sect.~\ref{sec:timing}). This is applied to known sources, which provides a higher sensitivity due to source-specific information such as the location, orbital parameters, and flux spectrum.

For each source, GAPP extracts the pulsed portion of the pulsar's signal (see Sect.~\ref{sec:timing}). However, the un-pulsed flux of a discrete source can be obtained by fitting the steps in count rates that occur when the source rises or sets over the
Earth's horizon (see the GBM Earth Occultation Method
--GEOM-- web page\footnote{\url{https://gammaray.nsstc.nasa.gov/gbm/science/earth_occ.html}.} and \citealt{Wilson-Hodge12}).


The GAPP also inherited data from previous missions like the Burst and Transient Source Experiment (BATSE; \citealt{Fishman92,Bildsten1997}) on board the Compton Gamma Ray Observatory (CGRO; \citealt{Gehrels94}). 
One of the larger transient monitors in recent history, BATSE comprised eight NaI(Tl) large area detectors (LAD) each with 2025 cm$^2$ of geometric area \citep{Fishman92}. A plastic charged particle anticoincidence detector was in front of each LAD, resulting in a lower energy threshold of $\sim20$ keV. BATSE also included eight spectroscopy detectors that were not used for pulsar monitoring. The BATSE data consisted of nearly continuous time-binned data DISCLA (four channels, 1.024 s resolution) and CONT (16 channels, 2.048 s resolution). Generally, the first BATSE DISCLA channel, 20-50 keV, was used for pulsar monitoring. Comparatively, GBM detectors only have a Beryllium window in front and so they can reach $\sim 8$ keV, much lower than the BATSE LADs could. Despite the larger area of the BATSE detectors, the sensitivity is similar for GBM and BATSE for detecting outbursts of XRPs, due to the added low-energy response of the GBM detectors along with the abundance of photons from the sources at those energies. 

The pointing strategies for GBM and CGRO differ, with Fermi operating in a sky-scanning mode and CGRO operating using inertial pointing. This resulted in the need to incorporate the detector response in an earlier step in the data analysis process for GBM to account for changing angular response as the spacecraft scanned the sky. Both missions were in low-Earth orbit, at a similar inclination and altitude, resulting in similar energy-dependent background rates. Both missions used the data when a source was visible above the Earth's horizon for XRP analysis, resulting in similar exposure times of $\sim 40$ ks per day per source (depending on the source declination). 

The GAPP (see Sect.~\ref{sec:timing}) is based on the technique developed for BATSE \citep{Finger+99,Wilson-Hodge99,Wilson+02,Wilson03}, which measured pulsed frequencies and pulsed flux for a number of XRPs. Similarly to GBM, BATSE measured the un-pulsed flux for these sources using Earth occultation \citep{Harmon04}. 

We have now consolidated available BATSE data within the GAPP web pages\footnote{\url{https://gammaray.nsstc.nasa.gov/gbm/science/pulsars.html}} so as to provide the community with 20 yr of pulsar monitoring spanning the last 30 yr.
We show that the combination of BATSE and GBM data, spanning over almost three decades, allows for the long-term study of XRPs that unveil otherwise unobservable phenomena.

\section{Timing}\label{sec:timing}

\subsection{Time corrections}\label{subsec:time_corr}

Before delving into the timing analysis of each detected pulsar, the epoch of observed events need to be corrected. All recorded epoch times, $t$, are barycentered to remove the effects of the satellite and the Earth's revolutions, thus correcting the times as if the reference system is located at the center of mass of the solar system. This correction process returns a final epoch time, $t'$, that takes into account the following contributions: the reference time $t_0$, clock corrections $\Delta_{\rm clock}$ (which account for differences between the observatory clocks and terrestrial time standards), the Roemer delay $\Delta_{\rm R}$ (which accounts for the classical light travel time across the Earth's orbit), the Einstein delay $\Delta_{\rm E}$ (which accounts for the time dilation from the moving pulsar, observatory, and the gravitational redshift caused by the Sun and planets or the binary companion), and the Shapiro delay $\Delta_{\rm S}$ (which represents the extra time required by the pulses to travel through the curved space-time containing the solar system masses). Combining these termsn in the equation for the final epoch, we have

\begin{equation*}
t' = t - t_0 + \Delta_{clock} + \Delta_{R}+ \Delta_{E}+ \Delta_{S}.
\end{equation*}

If the orbit of the pulsar is known, a further correction is applied to the pulse arrival times, a correction known as orbital demodulation. The pulsar emission time, $t^{\rm em}$, is computed from the Barycentric Dynamical Time (TDB) $t'$, as $t^{\rm em} = TDB - z$, where $z$ is the line-of-sight delay associated with the binary orbit of the pulsar \citep{Deeter+81,Hilditch01}:
\begin{equation}\label{eq:z_def}
z = a_x\,sin\,i\,[ sin\,\omega\,(cos\,E-e)+\sqrt{(1-e^2)}\,cos\,\omega\,sin\,E\, ].
\end{equation}

Here, $a_{\rm x}$ is the projected semi-major axis of the binary orbit, $i$ is the orbit's inclination relative to the plane of the sky, $w$ is the periastron angle, and $e$ is the binary orbit eccentricity, while $E$ is the eccentric anomaly as expressed in Kepler's equation
\begin{equation}
E - e\,sin\,E = \frac{2\pi}{P_{orb}}(t^{em} - \tau_p),
\end{equation}
where $P_{\rm orb}$ is the orbital period and $\tau_p$ is the periastron passage epoch.

\subsection{The phase model}
Pulsars represent excellent timing tools thanks to their very small moment of inertia, which allows precise measurements of the pulsar spin and spin derivative via a pulsar timing technique. This involves the regular monitoring of the rotation of the NS, by tracking the arrival times of individual observed pulses. For this, an average pulse profile is produced at any time to be used as a template, along with the assumption that any given observed profile is a phase-shifted and scaled version of the template.
This is encoded in the evolution of the pulse phase as a function of time ($\phi(t)$). This \textit{pulse phase model} can be represented as a Taylor expansion around the reference time $t_{\rm 0}$ as

\begin{equation}\label{eq:phasemodel}
\phi(t) = \phi_0 + \nu_0(t - t_0) + \frac{1}{2}\dot{\nu}(t-t_0)^2  + ...
\end{equation}

where $\nu_{\rm 0} = \nu(t=t_{\rm 0})$ (and $\phi_{\rm 0}= \phi(t=t_{\rm 0})$), while $\dot\nu$ is the pulse frequency derivative.
Pulsar timing deals with the determination of the pulse phase as accurately as possible in order to unambiguously establish the exact number of pulsar rotations between observations.

By fitting Eq.~\ref{eq:phasemodel} to the frequencies determined by the means of the power spectra or epoch-folding method, a preliminary phase model is estimated. This allows us to produce pulse profiles and, at the same time, to reduce the amount of data and computing time. In turn, pulse profiles are used to refine the phase model to a higher precision (i.e., by phase-connection, see Sec.~\ref{subsec:Fourier} and Sec.~\ref{subsec:f_fdot}). To extract the periodic signal, two main methods are considered in this work: 1) the harmonic expansion, and 2) the search in frequency and frequency derivative.

\subsection{Pulse profiles using a harmonic expansion}\label{subsec:Fourier}

The pulsar periodic signal can be represented by a Fourier harmonic series:

\begin{equation}\label{eq:countrate}
m_k = \sum^{N}_{h=1}a_h cos\left\{2\pi\,h\,\phi(t_k)\right\} + b_h sin\left\{2\pi\,h\,\phi(t_k)\right\},
\end{equation}

where $m_{\rm k}$ is the model count rate at time $t_{\rm k}$, $a_{\rm h}$ and $b_{\rm h}$ are the Fourier coefficients, $h$ is the harmonic number, $N$ is the number of harmonics, and $\phi(t_{\rm k})$ is the phase model.
Similar to BATSE \citep{Bildsten1997,Finger+99}, six harmonics are typically used to represent the pulse profile. This results in a reasonable representation of all observed sources' pulse profiles, while the employment of additional harmonics does not improve the pulse structure significantly.
To obtain the harmonic coefficients $a_h$ and $b_h$, a fit is performed to minimize the $\chi^2$ function:

\begin{equation}
\chi^2 = \sum^{M}_{k=1}\frac{\left\{x_k-(m_k+B_k)\right\}^2}{\sigma^2_{x_k}},
\end{equation}

where $x_{\rm k}$ and $\sigma_{\rm k}$ are the measured count rates and errors, respectively, $M=2N$ is the number of statistically independent points, and $B_{\rm k}$ is the background (the un-pulsed count rate level).
For this technique to work, a careful choice of the data length has to be considered. The interval needs to be short enough to guarantee that the phase model is not significantly changed between the beginning and the end of the observation, while at the same time, the interval needs to be long enough to include sufficient data, typically between 5 and 10 times the spin period of the measured source.

Following \citet{Bildsten1997} and\citet{Woods07}, the GBM pulsed flux (F$_{\rm pulsed}$) is obtained as the root mean squared (RMS) pulsed flux:

\begin{equation}
F_{pulsed} = \sqrt{\sum^{N}_{h=1} \frac{ a_h^2 + b_h^2 - (\sigma_{a_h} ^2 + \sigma_{b_h} ^2)}{2} }
\end{equation}

where 

\begin{equation}
\sigma_{a_h}^2 = \frac{4}{P^2}\sum^{P}_{i=1} \sigma_{r_i}^2 \cos^2{(2\pi\phi_i h)} \quad, \quad
\sigma_{b_h}^2 = \frac{4}{P^2}\sum^{P}_{i=1} \sigma_{r_i}^2 \sin^2{(2\pi\phi_i h)}
\end{equation}

and $P$ is the total number of phase bins, and $\sigma_{\rm r_{\rm i}}$ is the uncertainty in the count rate in the $i$th phase bin.
The spectral model used to combine the count rate of each source with the spectral response is an empirical model based on observations published in the literature.
This procedure ensures that the pulsed flux is unbiased against the energy dependence of pulsed flux commonly observed in accreting XRPs, because the F$_{\rm pulsed}$ is calculated in relatively narrow energy bins, and the effect of an incorrectly assumed spectral model has been calculated to affect the derived flux only marginally, i.e. at a $\sim5\%$ level \citep{Wilson-Hodge12}.

\subsection{Pulse profiles using a search in frequency and frequency derivative}\label{subsec:f_fdot}

Due to the spin evolution shown by accreting pulsars, it is useful to apply a technique that not only estimates the spin frequency but also its derivative. Consequently, a search over a grid of pulse frequencies and frequency derivatives is performed to find the best-fitting value.
The search range is often estimated using past measurements (depending on availability) otherwise, a safe interval of $\pm0.01\nu_{\rm 0}$ is used, where $\nu_{\rm 0}$ is the pulsar frequency estimated in Sect.~\ref{subsec:Fourier}).
For an estimation of the pulsar frequency derivative range, a maximum spin-up rate is obtained from accretion theory (e.g., \citealt{Parmar+89}) assuming canonical NS parameters,

\begin{equation}\label{eq:parmar}
\dot\nu = 1.9\times10^{-12}\mu^{2/7}_{30}L^{6/7}_{37}\, \rm Hz\,s^{-1},
\end{equation}

where $\mu_{\rm 30}$ is the magnetic moment of the NS in units of $10^{30}\,$G\,cm$^3$ and $L_{\rm 37}$ is the luminosity in units of $10^{37}\,$erg\,s$^{-1}$.
Typical spin-down rates are of the order of a few times $10^{-13}\,$Hz\,s$^{-1}$.
Eq.~\ref{eq:parmar} is considered applicable only on a limited range of relatively high-luminosity values \citep{Parmar+89}, a condition that is met for all analyzed sources when detected by GBM.

Once the frequency and frequency derivative search ranges are established, a grid of phase offsets from the phase model in Eq.~\ref{eq:phasemodel} is created

\begin{equation}
\delta\phi_k(\delta\nu,\dot\nu_q) = \delta\nu_p(\bar{t}_k-\tau)+\frac{1}{2}\dot\nu_q(\bar{t}_k-\tau)^2.
\end{equation}

where $\bar{t}_{\rm k}$ is the time at the midpoint of segment $k$, $\tau$ is a reference epoch chosen near the center of the considered time interval (that is, at the epoch $\tau$, $\delta\phi = 0$ by definition), $\delta\nu_p$ is an offset in pulse frequency from $\nu_{\rm 0}$ in Eq.~\ref{eq:countrate}, and $\dot\nu_{\rm q}$ its derivative.
Each offset in pulse phase leads to a a shift in the individual pulse profiles, applied as a modification of the estimated complex Fourier coefficient: 

\begin{equation}
\beta_{kh}(\delta\nu,\dot\nu_q) = (a_{kh}-i\,b_{kh})\,{\rm exp}\left\{-i\,2\pi\,h\,\delta\phi(\delta\nu,\dot\nu_q)\right\},
\end{equation}

where $a_{\rm kh}$ and $b_{\rm kh}$ are the harmonic coefficients for harmonic $h$ and profile $k$ from Eq.~\ref{eq:phasemodel}. The best frequency and frequency derivative within the search grid are determined using the $Y_n$ statistic, following \citet{Finger+99}.

\subsection{Phase offset estimation and model fitting}

Once pulse profile templates are obtained following the methods outlined in Sect.~\ref{subsec:Fourier} and Sect.~\ref{subsec:f_fdot}, a phase offset $\Delta\phi$ can be estimated by comparing the fitted pulse profiles with the obtained template. The $\Delta\phi$ and pulse amplitude $A$ of each pulse is then determined by fitting each pulse profile to the template ($T_h$) by minimization of

\begin{equation}
    \chi^2 = \sum^{M}_{k=1}\frac{|\alpha_{kh} - A\,T_h\,exp(-i\,2\pi\,h\,\Delta\phi_k)|^2}{\sigma^2_{kh}}.
\end{equation}{}

Here, $\alpha_{kh} = a_{kh} - i\,b_{kh}$ is the complex Fourier coefficient for harmonic $h$ and profile $k$, and $\sigma^2_{\rm kh}$ is the error on the real or imaginary component of $\alpha_{\rm kh}$.

Phase offsets are the signature that the observed spin frequency is modulated by some effect. If the offsets present a random, erratic behavior consistent with a constant value, then the phase model cannot be improved, and the offsets are considered noise. However, if the offsets show a (possibly periodic) pattern, the phase model can be improved by minimization of

\begin{equation}\label{eq:phase-mod}
    \chi^2 = \sum^{M}_{k=1}\frac{(\phi(t^{em}_k) - \phi^{model}(t^{em}_k))^2}{\sigma^2_{\phi(t^{em}_k)}},
\end{equation}{}

where $\phi(t^{\rm em}_{\rm k}))$ is the total measured pulse phase (the phase model used to fold the pulse profiles plus the measured offset), $\sigma^2_{\phi(t^{\rm em}_{\rm k})}$ is the error on $\phi(t^{\rm em}_{\rm k}))$, and $\phi^{model}(t^{\rm em}_{\rm k}))$ is the new phase model that has been used in the fit.
Typically, the Levenberg-Marquardt method \citep{Press92} is used for the minimization of Eq.~\ref{eq:phase-mod}.
Such a process constrains the pulsar binary orbit by considering $t^{\rm em}=TDB - z$, remembering that TDB is the Dynamical Barycentric Time and $z$ is the line-of-sight delay associated with the binary orbit (see Sect.~\ref{subsec:time_corr} and Eq.~\ref{eq:z_def}).


\section{Overview of Accreting X-Ray Pulsars}\label{sec:overview}

XRPs that are part of binary systems can be classified into two groups, according to the mass of the donor star \citep{Lewin97}:
\begin{itemize}
\item High Mass X-ray Binaries (HMXBs) are systems where the donor star is a massive O or B stellar type, typically with $M\geq5\,M_\odot$. 
The system is generally younger, and the stellar wind is strong.
When the compact object is a NS, its magnetic field is of the order of $10^{12}\,$G. 
In our Galaxy, these objects are mostly found on the Galactic plane, especially along the spiral arms.

\item Low Mass X-ray Binaries (LMXBs) are systems where the donor star is a spectral type A or later star, or a white dwarf with a mass of $M\leq1.2\,M_\odot$. These systems are generally older than HMXBs, with weaker stellar winds from the donor. The NS magnetic field observed in these systems has decayed to about $10^{7-8}\,$G. Moreover, LMXBs are typically found toward the Galactic center; although, some of them have been observed in globular clusters.

\end{itemize}

XRPs are largely found in HMXBs. In fact, GBM detected XRPs are almost exclusively HMXBs. Depending on the binary system properties, three different methods of mass transfer can take place in X-ray binaries:

\begin{itemize}
\item[1.] \textit{Wind-fed systems}: Accretion from stellar winds is particularly relevant when the donor star is a massive main-sequence or supergiant O/B star, because those winds are dense, with mass-loss rates of $\dot{M}_{\rm w}\approx 10^{-6}-10^{-7}\,M_\odot\,$yr$^{-1}$.
Typically, in wind-fed systems, the compact object orbits the donor star at a close distance, thus, being deeply embedded in the stellar wind, accreting at all orbital phases.
These systems are therefore \textit{persistent} sources, showing variability on a timescale that is much shorter than the orbital period (i.e., $10^2-10^4\,$s).

\item[2.] \textit{Roche lobe-overflow (RLO) systems}: When the binary system is such that the donor star radius is larger than its Roche lobe, the star loses part of its material through the first Lagrangian point $L_1$. When RLO takes place, the mass flow does not directly impact the compact object due to the intrinsic orbital angular momentum of the transferred material. Instead, it forms an accretion disk around the compact object. Since the transfer of matter is generally steady, RLO systems are also persistent sources.

\begin{figure*}[!t]
\includegraphics[width=1.\textwidth]{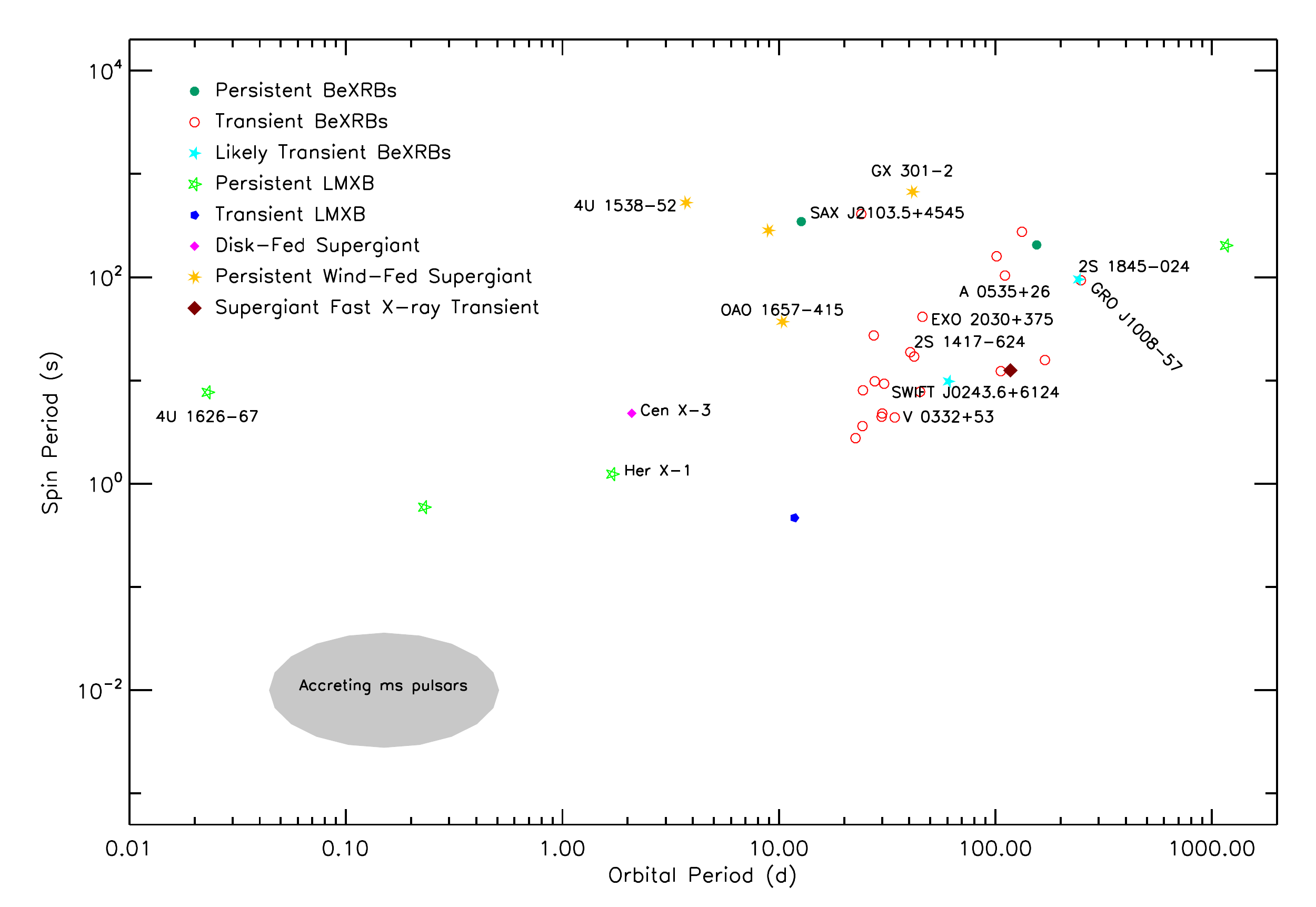}
\caption{The Corbet diagram showing spin period (y-axis) versus orbital period (x-axis) for all GBM detected accreting XRPs with known orbital period. A few representative sources have been labelled.
The region populated by accreting millisecond pulsars (grey oval) has also been labelled for comparison.}
\label{fig:corbet}
\end{figure*}

\item[3.] \textit{Be/X-Ray Binary systems (BeXRBs)}: In these systems, the donor star is an O- or B-type star that expels its wind on the equatorial plane under the form of a circumstellar disk (also called the Be disk). The disk is composed of ionized gas that produces emission lines (especially H$\alpha$).
When the orbiting compact object (CO) passes close to or through the Be disk, a large flow of matter is pulled from the disk, forming an accretion disk around the CO by its gravitational potential. Subsequently, matter is then accreted onto the CO giving rise to an X-ray outburst.
Due to the orbital modulation of the accreted matter, these systems show only \textit{transient} activity.
X-ray outbursts in BeXRBs are classified into two types:
\begin{itemize}
\item{Type I}: also called \textit{normal} outbursts. These are less luminous outbursts, with a peak luminosity of $\sim10^{36-37}\,$erg\,s$^{-1}$, occurring typically at periastron passages and lasting for a fraction of the orbital period.

\item{Type II}: also called \textit{giant} outbursts. These episodes are more rare, more luminous (peak luminosity of $\sim10^{37-38}\,$erg\,s$^{-1}$), and do not show any preferred orbital phase, lasting for a large fraction of the orbital period or even for several orbits.

\end{itemize}
\end{itemize}

Despite the aforementioned classifications, the zoo of XRPs often shows systems that have properties belonging to different classes and are characterized by mixed mass transfer modalities. For example, theoretical and observational works show that wind-captured disks can form around the CO of certain HMXBs (see, e.g., \citealt{Jenke2012,Blondin13,ElMellah19}).
Moreover, the recent discovery of new systems led to the classification of additional subclasses, e.g., the HMXBs with supergiant companions (SgXBs), and the Super-giant Fast X-ray Transients (SFXTs; \citealt[and references therein]{Sidoli18}).
However, different subclasses may also represent similar systems observed at different accretion regimes or at different evolutionary stages.
For example, gated accretion models are invoked to explain the variable activity of SFXTs, where the transitions between possible regimes are triggered by the inhomogeneous (i.e., clumpy) ambient wind \citep[and references therein]{Bozzo16, MartinezNunez17, Pradhan18}.

All GBM detected XRPs and relevant properties are summarized in Table~\ref{tab:summary}. The different classes for these sources are shown in Fig.~\ref{fig:corbet}, where they are plotted in the Corbet diagram \citep{Corbet1986}.

\begin{longrotatetable}
\begin{deluxetable}{lccccccccccc}
\tabletypesize{\scriptsize}
\tablecaption{GBM X-Ray Pulsars: Coordinates, Orbital Elements, Spin Periods and Distances.\label{tab:summary}}
\tablewidth{\textwidth}
\tablehead{
\colhead{Source} & \colhead{Class} & \colhead{R.A.} & \colhead{Decl.} & 
\colhead{$P_{\rm orb}$} & \colhead{$P_{\rm spin}$} & 
\colhead{$T_{\pi/2}^*$} & \colhead{$a_{\rm x}$sin\,$i$} & 
\colhead{$w$} & \colhead{$e$}  & \colhead{$d^{\dagger}$}\\
\colhead{} & \colhead{} & \colhead{($^\circ$)} & \colhead{($^\circ$)} & 
\colhead{(days)} & \colhead{(s)} &
\colhead{(MJD)} & \colhead{(l s)} & \colhead{($^\circ$)} & \colhead{} & \colhead{(kpc)} } 
{\startdata{
 &  &  & & & $Transient$   &   &  &  &  &   \\
GRO J1744--28\tablenotemark{a}    & LMXB-RLO  & 266.1379 & -28.7408 & 11.8358(5) & 0.467046314
& 56692.739(2) & 2.639(1) & 0.00 
 & $<6\times10^{-3}$ & $8.5^{+2.0}_{-4.5}$\, [\tablenotemark{b}]\\
SAX J2103.5+4545\tablenotemark{a} & BeXRB & 315.8988 & 45.7515 & $12.66528(51)$ & 358.61 & $52545.411(24)$ & $80.81(67)$ & $241.36(2.18)$ & $0.401(18)$ & $6.4^{+0.9}_{-0.7}$\\
4U 1901+03\tablenotemark{a}      & BeXRB & 285.9047 & 3.1920   & 22.5348(21) & 2.761792
& 58563.8361(8) & 106.989(15) & 268.812(3) & 0.0363(3) & $2.2^{+2.2}_{-1.3}$ \\ 
RX J0520.5--6932\tablenotemark{a} & BeXRB  & 80.1288 & -69.5319 & 23.93(7) & 8.037 & 56666.41(3) & 107.6(8) & 233.50 & 0.0286 & LMC \\ 
A 1118--615\tablenotemark{a}      & BeXRB & 170.2408 & -61.9161 & 24.0(4) &  407.6546
& 54845.37(10) & 54.8(1.4) & 310(30) & 0.10(2) & $2.93^{+0.26}_{-0.22}$\\ 
4U 0115+63\tablenotemark{a}      & BeXRB & 19.6329 & 63.7400 & 24.316895 & 3.61 & 57963.237(3) & 141.769(72) & 49.51(9) & 0.3395(2) & $7.2^{+1.5}_{-1.1}$ \\ 
Swift J0513.4--6547\tablenotemark{a} & BeXRB & 78.3580 & -65.7940 & 27.405(8) & 27.28 & 54899.02(27) & 191(13) & ... & $<0.17$ & LMC\\
Swift J0243.6+6124\tablenotemark{a} & BeXRB & 40.9180 & 61.4341 & 27.587(17) & 9.86  & 58103.129(17) & 115.84(32) & -73.56(16) & 0.09848(42) & $6.9^{+1.6}_{-1.2}$\\
GRO J1750--27\tablenotemark{a}    & BeXRB  & 267.3046 & -26.6437 & 29.803890 & 4.45 & 49931.02(1) & 101.8(5) & 206.3(3) & 0.360(2) & $18.0^{+4.0}_{-4.0}$ [\tablenotemark{b}]\\
Swift J005139.2--721704\tablenotemark{a} & BeXRB  &  12.9116 &  -72.284666 & 20-40 & 4.8 & ... & ...& ... & ... & SMC \\
2S 1553--542\tablenotemark{a}     & BeXRB & 239.4542 & -54.4150 & 31.34(1) & 9.29  & 57088.927(4) & 201.48(25) & 164.8(1.2) & 0.0376(9) &  $20.0^{+4.0}_{-4.0}$ [\tablenotemark{b}] \\ 
V 0332+53\tablenotemark{a}       & BeXRB & 53.7495 & 53.1732 & 33.850(3) & 4.37 & 57157.38(5)  & 77.8(2)  & 277.4(1) & 0.371(5) & $5.1^{+1.1}_{-0.8}$ \\
XTE J1859+083\tablenotemark{a} & BeXRB & 284.7700 & 8.2500 & 37.97 & 10.0 & 57078.7 & 57100.5(5) & 211.4(1.8) & -117.0(0.9) & $2.7^{+2.4}_{-1.5}$ \\
KS 1947+300\tablenotemark{a}     & BeXRB & 297.3979 & 30.2088 & 40.415(10) & 18.81 & 51985.31(7) & 137(3) & 33(3) & 0.033(13) & $15.2^{+3.7}_{-2.8}$ \\
2S 1417--624\tablenotemark{a}     & BeXRB & 215.3000 & -62.7000 & 42.19(1) & 17.51 & 51612.17(5) & 188(2) & 300.3(6) & 0.446(2)& $3.8^{+2.8}_{-1.8}$ \\
SMC X-3\tablenotemark{a}         & BeXRB & 13.0237 & -72.4347 & 45.04(8) & 7.81 & 57676.4(3) & 190.3(1.3) & 240.3(1.1) & 0.244(5) & SMC \\
EXO 2030+375\tablenotemark{a}    & BeXRB & 308.0633 & 37.6375 & 46.0213(3) & 41.33 & 52756.17(1) & 246(2) & 211.9(4) & 0.410(1) & $3.6^{+1.4}_{-0.9}$ \\
MXB 0656--072\tablenotemark{a}    & BeXRB & 104.6125 & -7.2633 & 101.2 & 160.7 & ... & ... & ... & $0.4$\tablenotemark{b} & $5.1^{+1.4}_{-1.0}$ \\
GS 0834--430\tnote{a}     & BeXRB & 128.9792 & -43.1850 & 105.8(4) & 12.3 & ... & ... & ... & (0.10-0.17) & $5.5^{+2.5}_{-1.7}$\\
GRO J2058+42\tablenotemark{a} & BeXRB & 314.6987 & 41.7743 & 110(3) & 193.61
& ... &... & ... & ...& $8.0^{+1.2}_{-1.0}$ \\
A 0535+26\tablenotemark{a}       & BeXRB  & 84.7274 & 26.3158 & 111.1(3) & 103.5 & 49156.7(1.0) & 267(13) & 130(5) & 0.42(2) & $2.13^{+0.26}_{-0.21}$ \\
IGR J19294+1816\tablenotemark{a} & BeXRB & 292.4829 & 18.3107 & 117.2 or 22.25\tablenotemark{b} & 12.45 & ... & ... & ... & ... & $2.9^{+2.5}_{-1.5}$  \\
GX 304--1\tablenotemark{a}        & BeXRB & 195.3213 & -61.6018 & 132.18900 & 272.0 & 55425.6(5) & 601(38) & 130(4) & 0.462(19) & $2.01^{+0.15}_{-0.13}$ \\
RX J0440.9+4431\tablenotemark{a} & BeXRB & 70.2472 & 44.5304 &150.0(2) & 202.5
& ... & ... & ... & $>0.4$\tablenotemark{b} & $3.2^{+0.7}_{-0.5}$  \\ 
XTE J1946+274\tablenotemark{a}   & BeXRB & 296.4140 & 27.3654 & 172.7(6) &  15.74974
& 55515.0(1.0)& 471.2(4.3) & -87.4(1.7) & 0.246(9) & $12.6^{+3.9}_{-2.9}$ \\
2S 1845--024\tablenotemark{a}     & BeXRB & 282.0738 & -2.4203  & 242.180(12) & 94.6 & 49616.48(12) & 689(38) & 252(9) & 0.8792(54)  & $10.0^{+2.5}_{-2.5}$ [\tablenotemark{b}]\\ 
GRO J1008--57\tablenotemark{a}    & BeXRB & 152.4420 & -58.2933 & 249.48(4) & 93.7134
& 54424.71(20) & 530(60) & -26(8) & 0.68(2) & $5.8^{+0.5}_{-0.5}$\, [\tablenotemark{b}] \\
Cep X-4\tablenotemark{a}         & BeXRB & 324.8780 & 56.9861 & (23-147) & 66.3  & ... & ... & ... & ... & $10.2^{+2.2}_{-1.6}$ \\
IGR J18179--1621\tablenotemark{a} & HMXB & 274.4675 & -16.3589 & -- & 11.82
& ... & ... & ... & ... & $8.0^{+2.0}_{-7.0}$ [\tablenotemark{b}] \\
MAXI J1409--619\tnote{a}  & BeXRB? & 212.0107 & -61.9834 & -- & 506
& ... & ... & ... & ... & $14.5^{+2.0}_{-2.0}$ [\tablenotemark{b}] \\
XTE J1858+034\tablenotemark{a}   & BeXRB & 284.6780 & 3.4390 & ... & 221.0  & ... & ... & ... & ... & $1.55^{+0.28}_{-0.21}$ \\
\hline
  &  &  &  &  & $Persistent$ &  &  &  &  &   \\
4U 1626--67\tablenotemark{a}    & LMXB - RLO  & 248.0700 & -67.4619 & 0.02917(3) & 7.66  & ... & ... & ... & ... & $3.5^{+2.3}_{-1.3}$ \\
Her X-1\tablenotemark{a}         & LMXB - RLO & 254.4571 & 35.3426 & 1.700167590(2) &1.237& 46359.871940(6)&13.1831(4)&96.0(10.0)& 4.2(8)E-4 & $5.0^{+0.8}_{-0.6}$ \\
Cen X-3\tablenotemark{a}         & sgHMXB - RLO + wind &170.3133&-60.6233&2.08704106(3)&4.8& 50506.788423(7) &39.6612(9)&--&$<0.0001$ & $6.4^{+1.4}_{-1.1}$ \\
4U 1538--52\tablenotemark{a}      & sgHMXB - wind & 235.5971 & -52.3861& 3.7284140(76) & 526.8& 52 855.061(13) & 53.1(1.5) & 40(12) & 0.17(1) & $6.6^{+2.2}_{-1.5}$ \\
Vela X-1\tablenotemark{a}        & sgHMXB - wind & 135.5286 & -40.5547 & 8.964427(12) & 83.2 & 42 611.349(13) & 113.89(13) & 152.59(92)& 0.0898(12) &$2.42^{+0.19}_{-0.17}$ \\
OAO 1657--415\tablenotemark{a}    & sgHMXB - wind & 255.2038 & -41.6560 & 10.447355(92) & 37.1 & 52674.1199(17) & 106.157(83) & 92.69(67) & 0.1075(12) & $7.1\pm1.3$ [\tablenotemark{b}]  \\
GX 301-2\tablenotemark{a}        & hgHMXB - wind & 186.6567 & -62.7703 & 41.506(3) & 684.1618 & 53532.15000 &368.3(3.7) & 310.4(1.4) & 0.462(14) & $3.5^{+0.6}_{-0.5}$ \\
GX 1+4\tablenotemark{a}          & LMXB & 263.0128 & -24.7456 & 1160.8(12.4) & 159.7 & 51942.5(53.0)  & 773(20) & 168(17) & 0.101(22) & $7.6^{+4.3}_{-2.8}$  \\ 
}\enddata}
\tablecomments{Sources are listed from top to bottom in order of increasing orbital period. $^*$Mid-eclipse time, equivalent to the time when the mean longitude $l=\pi/2$ for a circular orbit; $^{\dagger}$Distances obtained from the second Gaia Data Release (DR2) (unless specified otherwise); Large and Small Magellanic Clouds (LMC and SMC) are considered at 50 and 62 kpc, respectively. Spectral models, orbital parameters and distances for targets unavailable in the \textit{Gaia} DR2 are obtained for each source from the following works:}
  \begin{tablenotes}[para]
  \small\medskip
  \item[J1744a]{\citet{Sanna17};}
  \item[J1744b]{\citet{Nishiuchi99};}
  \item[J2103] {\citet{Camero07};}
  \item[4U1901] {\citet{Galloway+05,Jenke+Finger11}, and this work;}
  \item[J0520] {\citet{Kuehnel14};}
  \item[A1118] {\citet{Staubert11};}
  \item[4U0115] {This work;}
  \item[J0513] {\citet{Coe15};}
  \item[J0243] {\citet{Jenke18};}
  \item[J1750a] {\citet{Scott97} with period correction from GBM data;}
  \item[J1750b] {\citet{Lutovinov19};}
  \item[J005139] {\citet{Laycock2003};}
  \item[2S1553a] {This work;}
  \item[2S1553b] {\citet{Tsygankov16};}
  \item[V0332] {\citet{Doroshenko+16};}
  \item[J1859] {\citet{Kuehnel+16};}
  \item[KS1947] {\citet{Galloway+04};}
  \item[2S1417] {\citet{Finger+96,Inam+04};}
  \item[SMCX-3] {\citet{Townsend+17};}
  \item[EXO2030] {\citet{Wilson+08};}
  \item[MXB 0656a] {\citet{Morgan+03};}
  \item[MXB 0656b] {\citet{Yan12};}
  \item[GS0834] {\citet{Wilson97};}
  \item[J2058] {\citet{Wilson+98};}
  \item[A0535] {\citet{Finger0535};}
  \item[J19294a] {\citet{Corbet+Krimm09};}
  \item[J19294b] {\citet{Cusumano+16};}
  \item[GX304] {\citet{Sugizaki+15}
  }
  \item[J0440a] {\citet{Ferrigno+13};}
  \item[J0440b] {\citet{Yan+16};}
  \item[J1946a] {\citet{Marcu+15};}
  \item[J1946b] {\citet{Orlandini12};}
  \item[2S1845a] {\cite{Finger+99};}
  \item[2S1845b] {\cite{Koyama90};}
  \item[J1008a] {\citet{Coe+07,Kuehnel+13};}
  \item[J1008b] {\citet{Riquelme+12};}
  \item[CepX4] {\citet{Wilson99};}
  \item[J18179a] {\citet{Halpern12};}
  \item[J18179b] {\citet{Nowak12}; }
  \item[J1409a] {\citet{Kennea+10};}
  \item[J1409b] {\citet{Orlandini12};}
  \item[J1858] {\citet{Remillard98};}
  \item[4U1626] {\citet{Chakrabarty98};}
  \item[Her X-1] {\citet{Staubert+09}.}
  \item[Cen X-3] {\citet{Raichur+Paul10, Falanga+15};} 
  \item[4U1538] {\citet{Falanga+15,Clark00};} 
  \item[Vela X-1] {\citet{Bildsten1997, Kreykenbohm+08, Falanga+15};}
  \item[OAO1657a] {\citet{Jenke+12,Falanga+15};}
  \item[OAO1657b] {\citet{Audley06};}
  \item[GX301] {\citet{Sato+86, Koh+97, Doroshenko+09};}
  \item[GX1+4] {\citet{Hinkle+06}.}
  \end{tablenotes}
\end{deluxetable}
\end{longrotatetable}

\section{Individual Accreting X-Ray Pulsars observed by GBM}\label{sec:individuals}

There are 39 sources in total, 31 transient systems, and eight persistent systems, with frequency and pulsed flux histories available on the GAPP public website\footnote{\url{https://gammaray.nsstc.nasa.gov/gbm/science/pulsars.html\#}}. For each source, we link the corresponding GAPP web page for the reader's convenience.
The main properties of each source are listed in Table~\ref{tab:summary}, along with their distance values as measured either by the Gaia mission \citep{Bailer-Jones18} following the method described in Appendix~\ref{sec:gaiadist} or as otherwise specified.

\subsection{Transient Outbursts in BeXRB systems}

Most GBM detected XRPs are BeXRB systems. Among the transient systems, pulsations from 28 BeXRBs, 1 possible BeXRB, 1 HXMB (with no better subclassification), and 1 LMXB are observed with GBM. Below, we describe the main timing properties of each transient XRB detected by GBM.

\subsubsection{GRO J1744--28}

GRO J1744--28 is the fastest accreting X-ray pulsar, with a spin period of only $\sim44\,$ms, discovered with BATSE \citep{Finger+96}. This source is also known as the \emph{Bursting Pulsar}, due to the fact that it shows Type II-like bursting activity, usually attributed to thermonuclear burning, but it is possibly due to accretion processes in GRO J1744--28 \citep[and references therein]{Court18}. It is the only LMXB among the transient systems detected by GBM. It has an orbital period of about $12\,$days, and its distance is calculated as $\sim8.5\,$kpc in \citet{Kouveliotou96} and \citet{Nishiuchi99}, but it is challenged by the value of $\sim4\,$kpc obtained from studies of its near-infrared counterpart, a reddened K2 III giant star \citep{Gosling17, Wang07,Masetti14}.
On the other hand, the closest Gaia counterpart is located at $14.0\arcsec$ from the nominal source position and at a distance of $1.3^{+1.2}_{-0.5}\,$kpc.
The activity observed from GRO J1744--28 is limited to three episodes: the Type II outburst that led to its discovery in 1995 \citep{Kouveliotou96}, the outburst that occurred in 1997 \citep{Nishiuchi99}, and the last one in 2014, which followed almost two decades of quiescence (\citealt{Dai15,Sanna17}, and references therein).
The spin-up rate observed during the outbursts is of the order of $10^{-12}\,$Hz\,s$^{-1}$, while the secular spin-up trend shows an average rate of about $2\times10^{-13}\,$Hz\,s$^{-1}$ \citep{Sanna17}.
GBM also measuredGBM also measured an average value of the spin derivative of $\sim3\times10^{-12}\,$Hz\,s$^{-1}$

an average value of the spin derivative of $\sim3\times10^{-12}\,$Hz\,s$^{-1}$
during the 2014 outburst\footnote{\url{https://gammaray.nsstc.nasa.gov/gbm/science/pulsars/lightcurves/groj1744.html}.}. Comparisons with archival BATSE data show a marginal long-term spin-up trend with an average rate of $\dot{\nu}\sim1\times10^{-14}\,$Hz\,s$^{-1}$.

\subsubsection{SAX J2103.5+4545}

SAX J2103.5+4545 was discovered by BeppoSAX as a transient accreting pulsar with a spin period of $\sim360\,$s  \citep{Hulleman98}.
With an orbital period of $\sim$13~days, it is amongst the shortest known for a BeXRB~\citep{Baykal07}. The Gaia distance for this source is $6.4^{+0.9}_{-0.7}\,$kpc, consistent with the distance value obtained from optical observations of the B0 Ve companion star ($6.5\,$kpc; \citealt{Reig04,Reig10}).
Although SAX J2103.5+4545 has been classified as a BeXRB \citep{Reig04}, it does not follow the Corbet $P_{\rm orb}$--$P_{\rm spin}$ correlation, but it is located in the region of wind accretors (see Fig.~\ref{fig:corbet}). Since its discovery, numerous Type I and Type II outbursts have been observed \citep{Camero07}. Since then, SAX J2103.5+4545 has been showing a general spin-up trend at different rates\footnote{\url{https://gammaray.nsstc.nasa.gov/gbm/science/pulsars/lightcurves/saxj2103.html}.} but with an average value of $\dot{\nu}\approx10^{-12}\,$Hz\,s$^{-1}$ \citep{Camero07}. Those authors also observe a spin-up rate steeper than the expected power-law correlation, with a 6/7 index as reported in Equations~(\ref{eq:torque2}) and (\ref{eq:parmar}) (see Fig. 13 in their work). During outburst episodes, the measured spin-up rate is $\dot{\nu}=-2.6\times10^{-12}\,$Hz\,s$^{-1}$ \citep{Ducci08}. However, long ($\sim$yr) spin-down periods have also been observed between outbursts, with $\dot{\nu}=4.2\times10^{-14}\,$Hz\,s$^{-1}$ \citep{Ducci08}.

\subsubsection{4U 1901+03}

4U 1901+03 was first detected in X-rays by the Uhuru mission in 1970--1971 \citep{Forman76}. Afterwards, the source remained undetected until 2003, when it underwent a Type II outburst that lasted for about 5 months and during which pulsations were detected at a spin period of about $3\,$s \citep{Galloway+05}.
The orbital period is $\sim23\,$days \citep{Galloway+05,Jenke+Finger11}.
The optical companion stellar type was uncertain until recent measurements were obtained by \citet{McCollum19}, who proposed a B8/9 IV star, which is consistent with the X-ray timing analysis that favors a BeXRB nature \citep{Galloway+05}.
The Gaia measured distance is $2.2^{+2.2}_{-1.3}\,$kpc, much closer than the initially proposed distance of ${\sim10}\,$kpc \citep{Galloway+05}.
However, optical spectroscopy of the optical companion, together with the separation between the \textit{Gaia} measurement and the Chandra derived position for this source \citep{Halpern19}, led \citet{Strader19} to favor a distance $>12\,$kpc for this system.
After the Type II outburst in 2003, the source has remained mostly quiescent, showing moderate activity in 2011 December \citep{Jenke+Finger11,Sootome11}, when a weak flux increase was observed, accompanied by a spin-up trend. The spin-up observed during the Type II outburst in 2003 was $2.9\times10^{-11}\,$Hz\,s$^{-1}$ \citep{Galloway+05}. More recently however, the source underwent another Type II outburst \citep{Kennea19,Nakajima19}. The GBM spin-up average rate\footnote{\url{https://gammaray.nsstc.nasa.gov/gbm/science/pulsars/lightcurves/4u1901.html}.} measured during the 2019 outburst episode was $1.4\times10^{-11}\,$Hz\,s$^{-1}$, similar to that of the previous Type II outburst. GBM also observed the source slowly spinning down between outbursts at an average rate of $4.2\times10^{-13}\,$Hz\,s$^{-1}$.

\subsubsection{RX J0520.5--6932}

RX J0520.5--6932 was discovered with ROSAT \citep{Schmidtke94}. The only pulsations were detected two decades later, when a Swift/XRT survey of the LMC in 2013 revealed RX J0520.5--6932 to have undergone an X-ray outburst, and XMM-Newton observations found a spin period of about $8\,$s \citep{Vasilopoulos14}.
The orbital period is $\sim24\,$days \citep{Coe01,Kuehnel14}. The optical counterpart is an O9 Ve star \citep{Coe01}, and the source is located in the LMC ($\sim50\,$kpc). The outburst observed in 2013 was the the first and only one since its discovery \citep{Vasilopoulos13b}. During that episode, a strong spin-up trend was observed by GBM\footnote{\url{https://gammaray.msfc.nasa.gov/gbm/science/pulsars/lightcurves/rxj0520.html}.}, at a rate of $\dot{\nu}=3.5\times10^{-11}\,$Hz\,s$^{-1}$.


\subsubsection{A 1118--616}

X-ray pulsations with a period of $406.5\,$s were discovered in A 1118--616 by Ariel 5. Initially interpreted as the binary period \citep{Ives75}, it was later identified as the pulsar spin period \citep{Fabian75,Fabian76}.
The first determination of the orbital period of $24\,$days was obtained later by \citet{Staubert11}. The optical companion is an O9.5 IV-Ve star, Hen 3-640/Wray 793 \citep{Chevalier75}, and the Gaia measured distance for this system is $2.9^{+0.3}_{-0.2}\,$kpc (although, other works locate it at about $5.2\,$kpc; \citealt{Janot81,Riquelme+12}). Its outburst activity is sporadic, with only three major outbursts since its discovery (see \citealt{Suchy11}, and references therein). The average spin-up rate observed during accretion is of the order of $(2-4)\times10^{-13}\,$Hz\,s$^{-1}$ (see, e.g., \citealt{Coe1A94}, and the relevant GAPP web page\footnote{\url{https://gammaray.nsstc.nasa.gov/gbm/science/pulsars/lightcurves/a1118.html}.}), while the secular trend between outbursts is a spin-down rate of about $-9.1\times10^{-14}\,$Hz\,s$^{-1}$ \citep{Mangano09, Doroshenko10A}. After the 2011 outburst, the source entered a quiescent period that is still ongoing at the time of writing, remaining undetected with GBM.

\subsubsection{4U 0115+634}


Pulsations at $\sim4\,$s from 4U 0115+634 were discovered by SAS-3 in 1978 \citep{Cominsky78}. The orbital period is $24\,$days \citep{Rappaport78}, and the optical companion is  V635 Cas, a B0.2 Ve star. It has a \textit{Gaia} measured distance of $7.2^{+1.5}_{-1.1}\,$kpc, consistent with the approximate value of $\sim7\,$kpc inferred by \citet{Negueruela01a} and \citet{Riquelme+12}. 4U 0115+634 shows frequent outburst activity, with Type II outbursts observed as often as Type I outbursts, at a quasi-periodicity of $3-5\,$years \citep{Negueruela01a,Negueruela01b}. The general spin period evolution trend shows spin-down during quiescence, as well as between outbursts. However, rapid spin-up episodes are observed during Type II activity, $\dot{\nu}\sim2.3\times10^{-11}\,$Hz\,s$^{-1}$ \citep{Li12}. This resulted in a secular spin-up trend \citep{Boldin13}. However, the secular trend has recently inverted, and the source started to show long-term spin-down as observed by GBM\footnote{\url{https://gammaray.nsstc.nasa.gov/gbm/science/pulsars/lightcurves/4u0115.html}.}. 

\subsubsection{Swift J0513.4--6547}

Swift J0513.4--6547 was discovered by Swift during an outburst and identified as a pulsar with a spin period of $28\,$s in the same observation \citep{Krimm09}.
The outburst lasted for about 2 months, after which the source entered quiescence interrupted only by a moderate re-brightening in 2014, when it showed a luminosity of the order of $10^{36}\,$erg\,s$^{-1}$ \citep{Sturm14,Sahiner16}.
The system is located in the LMC, and the optical companion is a B1 Ve star \citep{Coe15}. 
The peak spin-up rate observed by GBM\footnote{\url{https://gammaray.nsstc.nasa.gov/gbm/science/pulsars/lightcurves/swiftj0513.html}.} during the 2009 outburst is about $3\times10^{-10}\,$Hz\,s$^{-1}$ \citep{Finger+Beklen09, Coe15}, while during the quiescent period between 2009 and 2014, the source was spinning down at an average rate of $-1.5\times10^{-12}\,$Hz\,s$^{-1}$~\citep{Sahiner16}.

\subsubsection{Swift J0243.6+6124}

Swift J0243.6+6124 is the newest discovered source in the present catalog and among the brightest. 
It was first discovered by Swift and then independently identified as a pulsar by Swift and GBM, with a spin period of about $10\,$s \citep{Jenke2017, Kennea17}. The orbital period is $\sim27\,$days \citep{Jenke18}, and the optical counterpart is a late Oe-type or early Be-type star \citep{Bikmaev17}, with a Gaia measured distance of $6.9^{+1.6}_{-1.2}\,$kpc.
Following its discovery, the source entered a Type II outburst that lasted for $\sim150\,$days, becoming the first known galactic Ultra-Luminous X-ray (ULX) pulsar, with a peak luminosity of about $2\times\,10^{39}\,$erg\,s$^{-1}$ \citep{Wilson18}. During the outburst episode, the source showed dramatic spin-up at a maximal rate of $\sim2\times10^{-10}\,$Hz\,s$^{-1}$ \citep{Doroshenko17}. After the Type II outburst, the source kept showing weaker X-ray activity for a few of the following periastron passages\footnote{\url{https://gammaray.nsstc.nasa.gov/gbm/science/pulsars/lightcurves/swiftj0243.html}.}. During these later passages, the spin-down rate of the source was about 100 times slower ($\sim-2\times10^{-12}\,$Hz\,s$^{-1}$), even at an accretion luminosity of a few $10^{36}\,$erg\,s$^{-1}$ \citep{Doroshenko19,Jaisawal19}.
Only during the last exhibited outburst did the source show a spin-up trend again at a rate comparable to the previously observed spin-up phase. Currently, the source remains quiescent.

\subsubsection{GRO J1750--27}

Pulsations at $4\,$s from GRO J1750--27 were observed by BATSE during the same outburst that led to its discovery \citep{Wilson95,Scott97}. The orbital period is about $30\,$days, and the system is located at a distance of $\sim18\,$kpc \citep{Scott97, Lutovinov19}, with no Gaia DR2 counterpart (but with a DR1 solution of $1.4^{+1.9}_{-0.5}\,$kpc). No optical counterpart has been identified yet due to the location of the system beyond the Galactic center. However, following the classification of \citet{Corbet1986}, \citet{Scott97} identified GRO J1750--27 as a BeRXB. GRO J1750--27 shows only sporadic outburst activity, with only three outbursts detected since its discovery and only one observed by GBM\footnote{\url{https://gammaray.msfc.nasa.gov/gbm/science/pulsars/lightcurves/groj1750.html}.}. Local spin-up trends during these outbursts have been observed at a rate of about $1.5\times10^{-11}\,$Hz\,s$^{-1}$ \citep{Shaw09, Lutovinov19}. The source does not show any appreciable spin derivative during quiescent periods.

\subsubsection{Swift J005139.2--721704}

Pulsations at about 4.8 s were first discovered from the SMC source XTE J0052--723 with RXTE \citep{Corbet01}.
This source has recently been identified as coincident with Swift J005139.2--721704 in the SMC \citep{Strohmayer+2018} and is listed on the GAPP website with this name\footnote{\url{https://gammaray.nsstc.nasa.gov/gbm/science/pulsars/lightcurves/swiftj005139.html}.}.
\citet{Laycock2003} identified the source as a BeXRB, inferring its orbital period as $\sim20-40\,$days based on its pulsation period and its possible location on the Corbet diagram.
Pulsations from this source were observed with GBM only once, during its recent outburst in 2018. This represented only the second outburst ever observed from this source \citep[and references therein]{Monageng19}. These authors reported the source to show unusual spin-down trends during accretion, which may be due to orbital modulation.

\subsubsection{2S 1553--542}

Pulsations at $\sim9\,$ s from 2S 1553--542 were discovered by SAS-3 \citep{Kelley82}. The orbital period is about $31\,$days \citep{Kelley83}, and
the optical companion has been identified as a B1-2V type star \citep{Lutovinov16}.
The closest counterpart measured by Gaia is located at an angular offset of $5\arcsec.5$ from the nominal source position, at a distance of $3.5^{+2.6}_{-1.5}\,$kpc. However, a distance of $20\pm4\,$kpc has been reported by \citet{Tsygankov16} based on the assumption of accretion-driven spin-up. 
Since its discovery, the source has exhibited three outbursts, all of which were Type II (see \citealt{Tsygankov16}, and references therein). This behavior is interpreted in terms of the low eccentricity ($e\sim0.035$) of the binary orbit \citep{Okazaki01}. Local spin-up rates during accretion episodes were measured as $2.9\times10^{-11}\,$Hz\,s$^{-1}$ for the 2008 outburst \citep{Pahari12}, and $8.7\times10^{-12}\,$Hz\,s$^{-1}$ for the 2015 outburst\footnote{\url{https://gammaray.msfc.nasa.gov/gbm/science/pulsars/lightcurves/2s1553.html}.}. The spin-down rate measured between these outbursts is about $-4.0\times10^{-13}\,$Hz\,s$^{-1}$ \citep{Tsygankov16}.

\subsubsection{V 0332+53}

Pulsations at $\sim4\,$s from V 0332+53 were detected by the EXOSAT satellite \citep{Stella85}. The same observations revealed a moderately eccentric orbit ($e\sim0.3$) and an orbital period of about $34\,$days. The optical companion is an O8-9 Ve star, BQ Cam \citep{Honeycutt85,Negueruela99}, and the system distance was first estimated to be $2.2-5.8\,$kpc \citep{Corbet86}. This was later increased to $6-9\,$kpc \citep{Negueruela99}. Both findings are consistent with a Gaia measured distance of $5.1^{+1.1}_{-0.8}\,$kpc.
Since its discovery, the source has shown four Type II outbursts, each one lasting for a few orbital periods and reaching peak luminosities of $\sim10^{38}\,$erg\,s$^{-1}$ (see \citealt{Doroshenko+16}, and references therein).
The spin-up rate measured during outburst episodes is $(2-3)\times10^{-12}\,$Hz\,s$^{-1}$
\citep{Raichur10}.
However, as outburst activity from this source is relatively rare, the net secular spin derivative trend shows a slow spin-down\footnote{\url{https://gammaray.nsstc.nasa.gov/gbm/science/pulsars/lightcurves/v0332.html}.}, $\sim-5\times10^{-14}\,$Hz\,s$^{-1}$.


\subsubsection{XTE J1859+083}

Pulsations from XTE J1859+083 at ${\sim}10\,$s were discovered with RXTE \citep{Marshall99}. An orbital period of 60.6 days was first proposed by \citet{Corbet09}, based on the separation of a few outbursts. However, analysis of a series of outbursts in 2015 led to a refined orbital solution with an orbital period of 37.9 days \citep{Kuehnel+16}.
No optical companion has been identified yet, but the source is considered a BeXRB due to its position on the Corbet diagram. The closest counterpart measured by Gaia is located at an angular offset of $17\arcsec.3$, at a distance of $2.7^{+2.4}_{-1.5}\,$kpc. In 2015, the source showed a new bright outburst \citep[and references therein]{Finger15}, during which GBM\footnote{\url{https://gammaray.msfc.nasa.gov/gbm/science/pulsars/lightcurves/xtej1859.html}.} measured a strong spin-up rate of $\dot{\nu}\sim1.6\times10^{-11}\,$Hz\,s$^{-1}$, similar to the rate observed in 1999 \citep{Corbet09}.


\subsubsection{KS 1947+300}

KS 1947+300 was first discovered with Mir-Kvant/TTM \citep{Borozdin90} and successively re-discovered with BATSE as the pulsating source GRO J1948+32 with a spin period of $\sim19\,$s \citep{Chakrabarty95}. These were later identified as the same source, KS 1947+300 \citep{Swank00}.
The orbital period is $42\,$days, while the binary orbit is almost circular, $e\sim0.03$ \citep{Galloway+04}. The Gaia measured distance is $15.2^{+3.7}_{-2.8}\,$kpc, approximately consistent with the distance of $\sim10\,$kpc measured by \citet{Negueruela03} and that of $10.4\pm0.9\,$kpc measured by \citealt{Riquelme+12}, who also derived the stellar type (B0V) of the optical companion. 
KS 1947+300 is the only known BeXRB with an almost circular orbit that shows both Type I and II outbursts. During these outbursts, the source shows a spin-up trend, with a rate measured for the 2013 Type II outburst of $(2-4)\times10^{-11}\,$Hz\,s$^{-1}$ \citep{Galloway+04,Ballhausen16,Epili16}, while the source is spinning down between outbursts at an average rate of $-8\times10^{-13}\,$Hz\,s$^{-1}$, as measured by GBM\footnote{\url{https://gammaray.msfc.nasa.gov/gbm/science/pulsars/lightcurves/ks1947.html}.}. 

\subsubsection{2S 1417--624}

Pulsations with a period of $17.6\,$s were discovered from 2S 1417--624 with SAS-3 observations in 1978 \citep{Apparao80,Kelley81}. This source shows both Type I and Type II outbursts, as well as decade-long quiescent periods.
The orbital period is $42\,$days \citep{Finger96}, with the optical counterpart identified as a B-type (most likely a Be-type) star located at a distance of $1.4-11.1\,$kp \citep{Grindlay84}, while the measured Gaia distance\footnote{Recently \citet{Ji+2019} adopted a different Gaia counterpart to the source, which has a distance of $9.9^{+3.1}_{-2.4}\,$kpc. This estimated distance is however inconsistent with the inferred distance of $\sim20\,$kpc calculated using accretion-driven torque models.} is $3.8^{+2.8}_{-1.8}\,$kpc. The secular slow spin-down trend observed during quiescence is overshadowed by the large spin-up induced during its Type II outbursts, observed to be\footnote{\url{https://gammaray.msfc.nasa.gov/gbm/science/pulsars/lightcurves/2s1417.html}}, $1.3\times10^{-12}\,$Hz\,s$^{-1}$ \citep{Raichur10}. Recently, 2S 1417--624 entered a new giant outburst episode at an orbital phase of $\sim0.30$, similar to the previous outburst in 2009 \citep{Gupta18,Nakajima18,Ji+2019}.

\subsubsection{SMC X-3}

SMC X-3 was discovered with SAS-3 as a bright source in the Small Magellanic Cloud \citep{Clark78}. However, it was not until 2004 that Chandra data analyzed by \citet{Edge04} recognized this source as an $\sim8\,$s pulsar found by \citet{Corbet04} with RXTE. The orbital period of the binary system is $45\,$days \citep{Townsend+17}, and the optical companion is a B1-1.5 IV-V star \citep{McBride08}. In 2016, the source underwent a Type II outburst that reached a super-Eddington bolometric peak luminosity of $2.5\times10^{39}\,$erg\,s$^{-1}$, making SMC X-3 a BeXRB system that is also a ULX source \citep{Townsend+17}.
During that episode\footnote{\url{https://gammaray.msfc.nasa.gov/gbm/science/pulsars/lightcurves/smcx3.html}}, the NS spun-up at an outstanding rate of $6.8\times10^{-11}\,$Hz\,s$^{-1}$ \citep{Townsend+17}. Conversely, the long-term (measured from 1998 to 2012) spin-down trend has an average rate that is about 500 times slower, $-1.4\times10^{-12}\,$Hz\,s$^{-1}$~\citep{Klus14}.

\subsubsection{EXO 2030+375}

\begin{figure}[!t]
\includegraphics[width=0.45\textwidth]{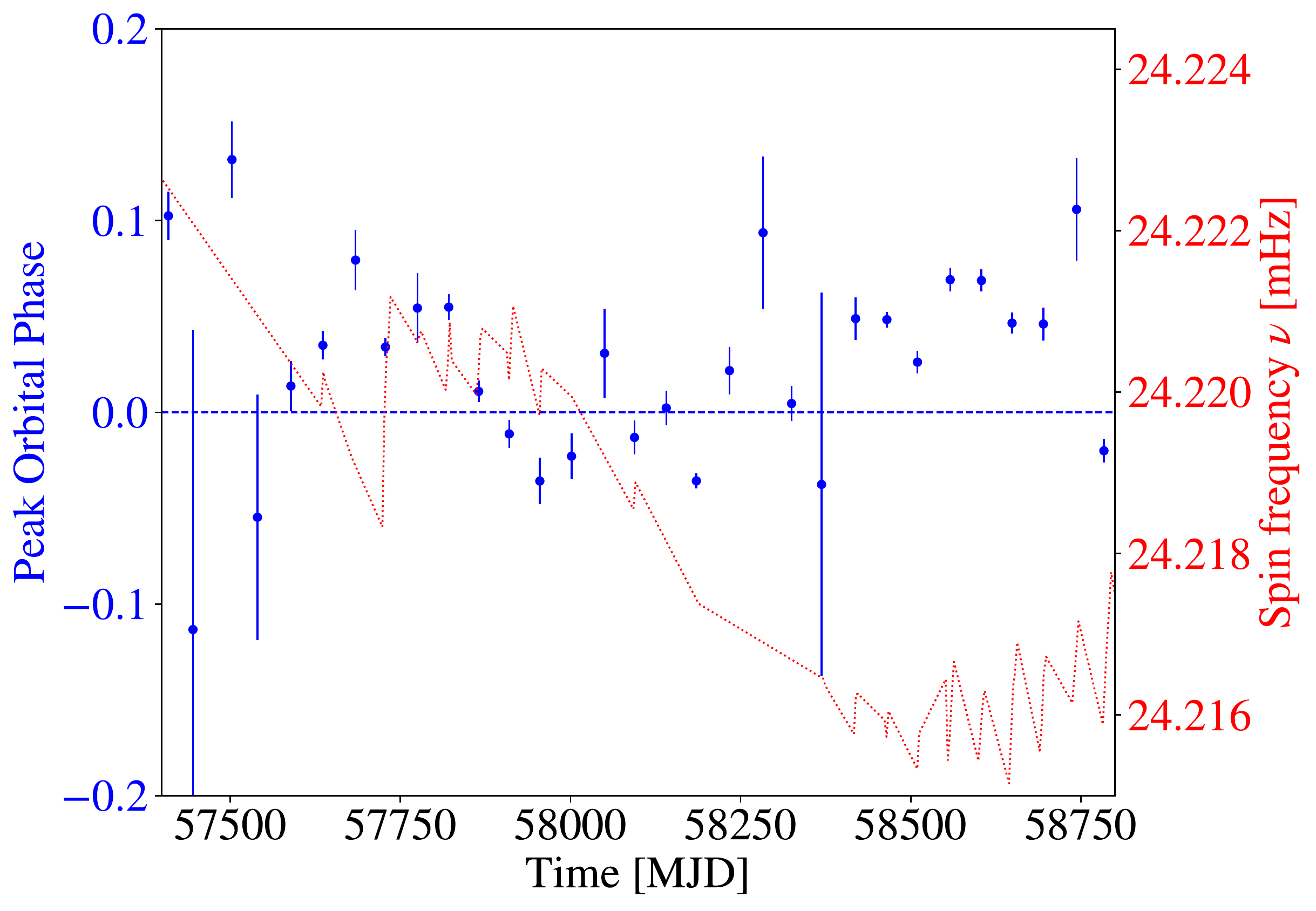}
\caption{Blue y-axis: The evolution of the orbital phase shift of Type I outbursts from EXO 2030+375 as measured with Swift/BAT (blue dots). Recently, the outbursts peak at $\sim0.1$ in orbital phase, a behavior that seems to be recurring with a periodicity of $\sim20\,$yr (see the text). ed y-axis: the evolution of the NS spin frequency (corrected for the orbital motion) as measured with GBM (red dashed line). The pulsar is now entering a new spin-up phase, after about $2000\,$days of spin-down, similar to what observed $\sim20\,$yr ago.}
\label{fig:phase_shifts}
\end{figure}

EXO 2030+375 is a transient source discovered with EXOSAT \citep{Parmar+89}.
The NS spin period is $41.7\,$s, while the orbital period is $\sim46\,$days \citep{Wilson+08}. The orbit of the NS around the O9-B2 stellar companion is eccentric, $e=0.4190$. Since its discovery, the source has shown both Type I and II outbursts. Type I episodes have been occurring nearly every orbit for $\sim28\,$yr with a typical duration of about $7-14\,$days, while Type II outbursts can last as long as $80\,$days. EXO 2030+375 is the XRP with the largest number of observed Type I outbursts ($\sim150$), detected in the X-ray band by many space-based observatories, e.g., Tenma, Ginga, ASCA, BATSE, RXTE, and more recently with Swift/BAT and GBM \citep{Laplace+17}. The long-term spin derivative trend observed by both BATSE and GBM\footnote{\url{https://gammaray.msfc.nasa.gov/gbm/science/pulsars/lightcurves/exo2030.html}.} is spin-up at a mean rate of $\dot{\nu}\sim1.3\times10^{-13}\,$Hz\,s$^{-1}$. During such long-term spin-up periods, outbursts occur typically $5-6\,$days after periastron passage. However, the source has shown two torque reversals, one of which was ongoing in 2019.
The first torque reversal occurred in 1995, which was first preceded by a ${\sim}3\,$yr quiescent period, accompanied later by a shift in the outburst peak to $8-9$ days earlier than the preceding outbursts ($3-4\,$days before periastron; \citealt{Reig+Coe98,Wilson+02}).
Recently, the source has shown another quiescent period ($\sim1\,$yr), after which the resumed activity was characterized by similar properties to those observed $\sim20\,$yr before - a shift in the outburst peak orbital phase and a spin-down trend. This behavior highlights a possible $21\,$yr cycle due to Kozai--Lidov oscillations in the Be disk \citep{Laplace+17}.
According to \citet{Laplace+17}, the shift in the peak orbital phase is $\sim0.15$ over the past cycle. To verify their predictions, we calculated the orbital shift of the outburst peak using the Swift/BAT monitor. To achieve this, we modeled each Type I outburst observed by the BAT with a skewed Gaussian profile, whose peak was taken to the corresponding outburst peak time.
These are shown in Fig.~\ref{fig:phase_shifts} as a function of time from 2016 January (MJD 57400) up to 2019 October (MJD 58700). At the time of writing, GBM recorded the start of a new spin-up phase, similar to what was observed in the previous cycle \citep{Laplace+17}. This supports the hypothesis formulated by those authors about a $\sim20\,$yr periodicity in the X-ray behavior of EXO 2030+375.

\subsubsection{MXB 0656--072}

Despite the discovery of MXB 0656--072 more than $40\,$yr ago with SAS-3 \citep{Clark75}, it took almost $30\,$yr years to detect any pulsations from this source with RXTE. RXTE observed the source to have a spin period of $\sim160\,$s \citep{Morgan+03}.
The orbital period is about $100\,$days \citep{Yan12}, and the optical companion is an O9.7 Ve star \citep{Pakull03, Nespoli12}. The Gaia measured distance is $5.1^{+1.4}_{-1.0}\,$kpc, consistent with the distance derived from optical analysis of the companion spectrum \citep{McBride06}. So far, the source has shown only Type I outbursts, with a peak luminosity of $<10^{37}\,$erg\,s$^{-1}$. The source has also shown fast spin-up during accretion. A spin-up trend of $\dot{\nu}\sim5\times10^{-12}\,$Hz\,s$^{-1}$ (that is about $0.45\,$s in $30\,$days) was observed in the 2003 outburst \citep{McBride06}. The last series of Type I outbursts observed from this source dates back to the period between $2007$ and $2009$ \citep{Yan12}; afterwards, the source entered a quiescent period that is still presently ongoing. The spin-up rate measured by GBM\footnote{\url{https://gammaray.msfc.nasa.gov/gbm/science/pulsars/lightcurves/mxb0656.html}} during the last of those outbursts was comparable to that measured in 2003.

\subsubsection{GS 0834--430}

Pulsations from GS 0834--430 were first observed with Ginga, each lasting $12\,$s \citep{Aoki92}. 
The orbital period was measured to be $106\,$days by \citet{Wilson97}. This was determined by using the spacing between the first five of seven outbursts observed between 1991 and 1993, while the last two were spaced by about $140\,$days.
The optical counterpart is a B0-2 III-Ve type star and estimated to be located at a distance of $3-5\,$kpc, inferred from the luminosity type \citep{Israel00}. This was later found to be consistent with the measured Gaia distance of $5.5^{+2.5}_{-1.7}\,$kpc for the closest counterpart located at $5.4\arcsec$ from the nominal source position.

The average spin-up rate during the first outbursting period was about $6\times10^{-12}\,$Hz\,s$^{-1}$ \citep{Wilson97}, while the spin-up rate measured by GBM\footnote{\url{https://gammaray.msfc.nasa.gov/gbm/science/pulsars/lightcurves/gs0834.html}} during the last outburst in 2012 was found to be $1.1\times10^{-11}\,$Hz\,s$^{-1}$ \citep{Jenke12}.

\subsubsection{GRO J2058+42}

Pulsations from this source were discovered by BATSE to have a spin period of 198 s during a giant X-ray outburst in 1996 \citep{Wilson96}. Subsequent observations of the source found an orbital period of about $110\,$days \citep{Wilson+98} and were consequently identified later as a BeXRB system \citep{Wilson+05}.
The first estimation of the distance to the source \citep{Wilson+98} was found to be $7-16\,$kpc away, consistent with the GAIA distance of $8.0_{-1.0}^{+1.2}\,$kpc.
During the giant outburst in 1996, the source showed spin-up at a rate of $1.7\times10^{-11}\,$Hz\,s$^{-1}$ \citep{Wilson+98}.
GRO J2058+42 was observed by GBM\footnote{\url{https://gammaray.msfc.nasa.gov/gbm/science/pulsars/lightcurves/groj2058.html}.} only during the recent bright X-ray outburst \citep{Malacaria19}, when the source showed a spin-up rate similar to that reported in 1996. It was previously unobserved by GBM, thus, representing the most recent addition to the GBM Pulsar catalog.

\subsubsection{A 0535+26}\label{subsec:a0535}

Pulsations from A 0535+26 were discovered by Ariel 5 with a period of $103\,$s \citep{Coe+75,Rosenberg+75}. The system has an orbital period of $111\,$days \citep{Nagase+82}. The optical counterpart is  HD 245770, an O9.7-B0 IIIe star located at a distance of $\sim2\,$kpc \citep{Hutchings+78,Li+79,Giangrande+80, Steele+98}. This distance was later confirmed by Gaia to be $2.1^{+0.3}_{-0.2}\,$kpc. The system regularly shows Type I outbursts, separated by both quiescent phases and Type II episodes (see \citealt{Motch+91, Mueller+13}, and references therein). Similar to GX~$1+4$ and GRO J1008--57 (see Sections~\ref{subsubsec:gx1+4} and \ref{subsec:groj10}, respectively), little to no spin-up is detected during Type I outbursts of this source. However, large spin-up trends have been observed during Type II outbursts. The source is found to be spinning down during quiescence\footnote{\url{https://gammaray.msfc.nasa.gov/gbm/science/pulsars/lightcurves/a0535.html}}. The average spin-down rate is $1.4\times10^{-11}\,$Hz\,s$^{-1}$ (see, e.g., \citealt{Hill+07}), while the measured spin-up rate during the giant outburst episodes is $\sim(6-12)\times10^{-12}\,$Hz\,s$^{-1}$ \citep{Camero+12,Sartore+15}.


\subsubsection{IGR J19294+1816}

IGR J19294+1816 was initially discovered with International Gamma-Ray Astrophysics Laboratory (INTEGRAL; \citealt{Turler09}), and later recognized as a pulsating source by Swift \citep{Rodriguez09a,Rodriguez09b}, with a pulsation period of $\sim12\,$s. An orbital period of $117$ days has been proposed, although this remains uncertain \citep{Corbet+Krimm09,Rodriguez09b,Bozzo11}. The Gaia measured distance is $2.9^{+2.5}_{-1.5}\,$kpc. However, independent measurements of the distance report inconsistent values. A lower limit has been estimated to be ${>}8\,$kpc for a B3 I optical counterpart by \citet{Rodriguez09a}, while a distance of $11\,$kpc was inferred by \citet{RR18} for a B1 Ve counterpart. Inspection of the GBM data\footnote{\url{https://gammaray.msfc.nasa.gov/gbm/science/pulsars/lightcurves/igrj19294.html}} reveals a secular spin-down trend with $\dot{\nu}\sim-2\times10^{-12}\,$Hz\,s$^{-1}$, interrupted by local spin-up episodes accompanying accretion during outbursts, with an average spin-up of about $\dot{\nu}\sim2.5\times10^{-11}\,$Hz\,s$^{-1}$. 
GBM observations support the $\sim$117\,days periodicity.
According to the most recent observations by GBM and \textit{XMM} \citep{Domcek19}, the source still exhibits a long-term spin-down trend, with Type I outbursts at each periastron passage.

\subsubsection{GX 304-1}\label{subsec:gx304}

Pulsations with a period of $\sim272\,$s from GX 304-1 were discovered with SAS-3 in 1978 \citep{McClintock77}. The orbital period is about $132\,$days, and the optical companion has been identified as a B2 Vne type star. The distance measured by Gaia to the companion, was found to be $2.01^{+0.15}_{-0.13}\,$kpc, in agreement with a previously measured distance of $2.5\,$kpc \citep{Mason78,Parkes+80}.
The source typically shows both Type I and II outbursts, as well as long ($\sim$yr) quiescent periods. According to GBM observations\footnote{\url{https://gammaray.msfc.nasa.gov/gbm/science/pulsars/lightcurves/gx304m1.html}}, the source shows accretion-driven spin-up episodes at a rate of about $\dot{\nu}\sim1\times10^{-12}\,$Hz\,s$^{-1}$ during active periods and long-term spin-up trends at an average rate of about  $\dot{\nu}\sim1.3\times10^{-13}\,$Hz\,s$^{-1}$. A spin-down rate between outbursts of $\dot{\nu}\sim-5\times10^{-14}\,$Hz\,s$^{-1}$ \citep{Malacaria+15,Sugizaki+15} has also been observed in the data. Recently, the source has entered a new period of quiescence, probably due to major disruptions of the Be disk following a Type II outburst, showing only sporadic X-ray activity \citep{Malacaria+17}.

\subsubsection{RX J0440.9+4431}

Pulsations with a period of $202\,$s from RX J0440.9+4431 were discovered with RXTE \citep{Reig99}. The binary orbital period is $150\,$days (\citealt{Ferrigno+13}, and references therein), and the optical companion is a B0.2 Ve star with a Gaia measured distance of $3.2^{+0.7}_{-0.5}\,$kpc, consistent with previous measurements of $\sim3.3\,$kpc \citep{Reig05}.
Only Type I outbursts have been observed from this source, the first and brightest of which was detected in 2010 \citep{Usui12}, exhibiting a spin-up rate of about $\dot{\nu}\sim4.5\times10^{-12}\,$Hz\,s$^{-1}$ in GBM\footnote{\url{https://gammaray.msfc.nasa.gov/gbm/science/pulsars/lightcurves/rxj0440.html}.}.
A strong, long-term spin-down trend has also been observed between the first pulsations discovered from the source in 1999 ($\approx206\,$s) and the outburst analyzed $12\,$yr later. Pulsation periods from the latter were measured at $\approx202\,$s, resulting in a spin-down rate of about $\dot{\nu}=-3\times10^{-12}\,$Hz\,s$^{-1}$ \citep{Ferrigno+13}.

\subsubsection{XTE J1946+274}

Pulsations with a period of $\sim15\,$s were first detected from XTE J1946+274 by RXTE \citep{Smith98}. The orbital period is about $169\,$days \citep{Wilson03}, and the optical companion is a B0-1 IV-Ve star with a measured Gaia distance of $12.6^{+3.9}_{-2.9}\,$kpc, consistent with previous measurements of $8-10\,$kpc (\citealt{Verrecchia02,Wilson03,Riquelme+12}).
A number of Type I and II outbursts have been observed from this source, as well as long quiescent periods\footnote{\url{https://gammaray.msfc.nasa.gov/gbm/science/pulsars/lightcurves/xtej1946.html}}. During accretion, the source shows strong spin-up at an average rate of $\dot{\nu}\sim(5-10)\times10^{-12}\,$Hz\,s$^{-1}$ \citep{Wilson03,Doroshenko17}. During quiescent periods, the source spins-down with a rate of about $\dot{\nu}=-2.3\times10^{-13}\,$Hz\,s$^{-1}$ over a long-term trend.
Recently, the source has shown another bright outburst episode, monitored with GBM and NICER \citep{Jenke18_J1946}. Analysis of these data is ongoing (Mailyan B. et al. 2020, in preparation).

\begin{figure*}[!t]
\includegraphics[width=1.\textwidth]{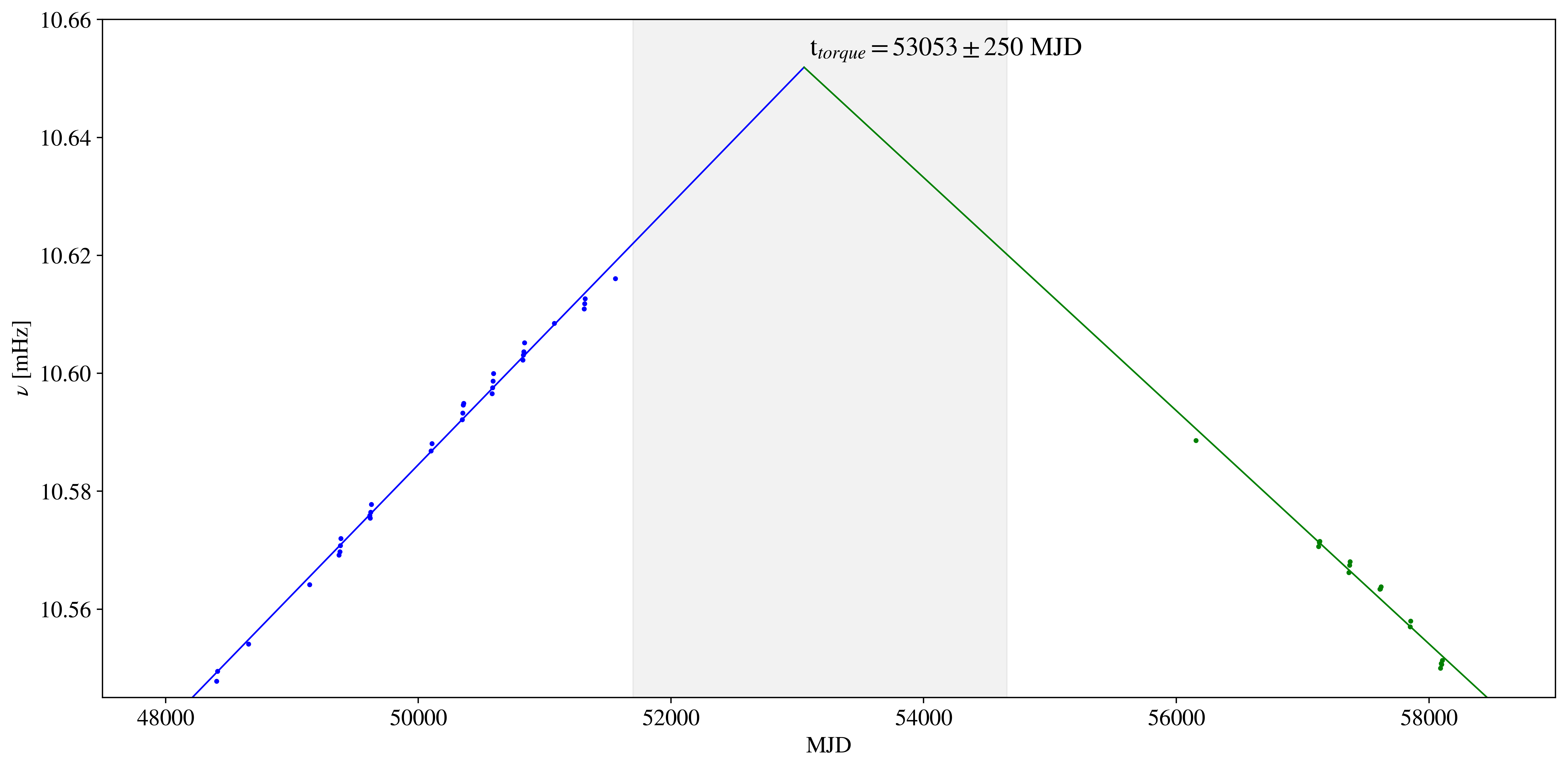}
\caption{The spin history of the BeXRB 2S 1845--024 as observed by BATSE (blue) and GBM (green). Errors are smaller than data points. Two separated linear fits are shown as straight lines for the BATSE (blue) and GBM (green) data. The estimated torque time is inferred from the intersection of the linear fit lines, $53053\pm250$ MJD (see Sect.~\ref{subsec:2s1845}). The grea shaded area marks the period where neither BATSE nor GBM data are available.}
\label{fig:2S1845_torque}
\end{figure*}

\subsubsection{2S 1845--024}\label{subsec:2s1845}

Pulsations from 2S 1845--024 (GS 1843--024) were discovered with Ginga with a spin period of $\sim30\,$s \citep{Makino88}. The orbital period is about $242\,$days \citep{Zhang96,Finger+99}. No optical counterpart is currently known for this system. However, based on the Corbet diagram and the regularity of the observed outbursts, this source has been identified as a BeXRB. No Gaia measurement of the distance is available for this source in the DR2, but an inferred distance of $\sim10\,$kpc has been obtained from the analysis of the X-ray spectral properties of the source \citep{Koyama90}. No Type II outbursts have been observed from this source.
A secular spin-up trend has been measured with BATSE during the first $\sim5.5\,$yr after its discovery. It results from fast local spin-up occurring during outburst episodes. This was found to occur at a rate of $\dot{\nu}\sim4\times10^{-12}\,$Hz\,s$^{-1}$, which yielded a long-term spin-up trend with a rate of $\dot{\nu}\sim2.7\times10^{-13}\,$Hz\,s$^{-1}$ \citep{Finger+99}. More recently, the source has inverted its long-term trend and has now been in a spin-down phase for around $10\,$yr.\footnote{\url{https://gammaray.nsstc.nasa.gov/gbm/science/pulsars/lightcurves/2s1845.html}} The strength of the local spin-up episodes associated with outbursting episodes ($\dot{\nu}\sim3.5\times10^{-12}\,$Hz\,s$^{-1}$), as well as that of the long-term spin-down trends ($\dot{\nu}\sim-2.4\times10^{-13}\,$Hz\,s$^{-1}$), is similar to the strength of those preceding the torque reversal. A comprehensive spin history for this source is shown in Fig.~\ref{fig:2S1845_torque}. 
An estimation of the torque reversal time can be inferred assuming that the long-term linear trends seen separately for BATSE data up to 51560 MJD and after 56154 MJD for GBM data,can be extrapolated to periods where neither BATSE nor GBM data were available. This returns a torque reversal time of $53053\pm250$ MJD, where the uncertainty is derived by extrapolating the two separate linear fits within the uncertainty of their parameters.


\subsubsection{GRO J1008--57}\label{subsec:groj10}

Pulsations from GRO~J1008--57 were discovered by CGRO during an X-ray outburst in 1993 \citep{Stollberg1993}. The NS has a spin period of about $93.5\,$s, while the binary orbital period is $\sim248\,$days \citep{Levine+06, Coe+07,Kuehnel+13}.
The optical counterpart is a either a dwarf (luminosity class III) or a supergiant (V) O9e-B1e type star \citep{Coe+94}. There is no available Gaia distance for this source, but \citet{Riquelme+12} estimates the system to be at a distance of either $9.7$ or $5.8\,$kpc, according to the luminosity type of the companion star.
As with GX~$1+4$ (see Sect.~\ref{subsubsec:gx1+4}) and A0535+26 (see Sect.~\ref{subsec:a0535}), the source exhibits a secular spin-down trend interrupted by brief spin-up episodes correlated with bright flux levels, typical of Type II outbursts\footnote{\url{https://gammaray.msfc.nasa.gov/gbm/science/pulsars/lightcurves/groj1008.html}}. The spin-up rate observed during the 2012 giant outburst is $6\times10^{-12}\,$Hz\,s$^{-1}$, while the secular spin-down rate is about $-2.3\times10^{-14}\,$Hz\,s$^{-1}$, induced by the propeller accretion mechanism in that regime \citep{Kuehnel+13}. The source typically undergoes an outburst at each periastron passage, with recent activity characterized by peculiar outburst light curves with $2-3$ peaks and a peak luminosity of several times $10^{37}\,$erg\,s$^{-1}$ \citep{Nakajima14, Kuhnel17}. Applying the orbital solution found for this source by \citet{Kuehnel+13} still shows orbital signatures in GBM data, and we therefore do not consider its pulse frequency history as demodulated.

\subsubsection{Cep X-4}

Pulsations from Cep X-4 at $\sim66\,$s were first detected with Ginga \citep{Makino+88}. Only a handful of outbursts with a relatively low luminosity have been observed from this source; thus, the orbital elements for this binary system are still unknown. However, a possible orbital period of about $21\,$days has been suggested by \citet{Wilson99} and \citet{McBride07}. The optical counterpart has been identified as a possible B1-2 Ve star \citep{Bonnet98}. The same authors have tentatively estimated a distance to the source of about $4\,$kpc, but this value has been challenged by \citet{Riquelme+12} who proposed a distance of either $7.9$ or $5.9\,$kpc according to whether the stellar type of the companion is a B1 or B2 star, respectively. The distance of $4\,$kpc also does not agree with the measured Gaia distance of $10.2^{+2.2}_{-1.6}\,$kpc.
Between 1993 and 1997, \citet{Wilson99} used BATSE data to measure an average spin-down rate of $\dot{\nu}\sim-4\times10^{-14}\,$Hz\,s$^{-1}$. Spin-up has also been observed during accretion episodes at a rate of $\sim10^{-12}\,$Hz\,s$^{-1}$.
At the time of writing, the source is still showing a general spin-down trend\footnote{\url{https://gammaray.msfc.nasa.gov/gbm/science/pulsars/lightcurves/cepx4.html}}.

\subsubsection{IGR J18179--1621}

Pulsations from IGR J18179--1621 were discovered by Swift with a spin period of about $12\,$s \citep{Halpern12}. This was later confirmed by Fermi/GBM \citep{Finger12} detections of its pulsations\footnote{\url{https://gammaray.msfc.nasa.gov/gbm/science/pulsars/lightcurves/igrj18179.html}}. The only activity reported from this source is the same that led to its discovery in 2012, when the source slightly brightened and became detectable by INTEGRAL and other X-ray satellites (see \citealt{Bozzo12}, and references therein). The nature of the optical companion is uncertain, but the analysis of the spectral characteristics of the source along with the presence of pulsations suggest that it belongs to the class of HMXBs/BeXRBs \citep{Nowak12,Tuerler12}. There is no measured Gaia distance for this source in the DR2, but a value of $8.0^{+2.0}_{-7.0}\,$kpc was found by \citet{Nowak12}.

\subsubsection{MAXI J1409--619}

Pulsations from MAXI J1409--619 were discovered by Swift, with an NS spin period of about $500\,$s \citep{Kennea+10} and later confirmed by GBM \citep{CameroAtel10}, which also detected a spin-up during a follow-up observation, as the source re-brightened a few weeks after its discovery. In the GBM data\footnote{\url{https://gammaray.msfc.nasa.gov/gbm/science/pulsars/lightcurves/maxij1409.html}}, the frequency increased at a rate of $1.6\times10^{-11}\,$Hz\,s$^{-1}$ during the outburst observed in 2010 December. Only a handful of observations have been carried out for this source immediately following its discovery. Consequently, very little is known about it. Given the shape of its light curve, the characteristics of its X-ray spectrum, its location close to the Galactic plane, and its vicinity of an infrared counterpart (2MASS 14080271--6159020), MAXI J1409--619 has been suggested to be an SFXT candidate \citep{Kennea+10}. There is no Gaia counterpart consistent with the X-ray source position as measured by Swift/XRT, but the distance to this source has been measured by \citet{Orlandini12} as $14.5\,$kpc. After the last GBM observation in 2010 December, the source has remained in a state of quiescence.

\subsubsection{XTE J1858+034}

Pulsations at a spin period of about $221\,$s from XTE J1858+034 were discovered by RXTE \citep{Remillard98, Takeshima98}. The source was discovered during one of only a few recorded outburst episodes (see also \citealt{Molkov04}), the last one being recorded by GBM in $2010$\footnote{\url{https://gammaray.nsstc.nasa.gov/gbm/science/pulsars/lightcurves/xtej1858.html}} \citep{Krimm10J1858}.
The orbital period of the binary is currently unknown, and the spectral type of the companion is still uncertain, but there are indications that it is a Be-type star \citep{Reig04atel,Reig05}. The closest counterpart measured by Gaia is located at an angular offset $3.5$\arcsec\, from the nominal source position, at a distance of $1.55^{+0.28}_{-0.21}\,$kpc.
There are no other available counterparts for this source.
The spin-up measured by GBM (uncorrected for binary modulation) during the 2010 accretion episode is about $1\times10^{-11}\,$Hz\,s$^{-1}$.

\subsection{Persistent binary systems}

\subsubsection{4U 1626--67}

4U~1626--67 is an LMXB discovered by Uhuru in 1977 \citep{Giacconi+72}.
The NS spins with a period of $7.66\,$s while orbiting its companion star, \emph{KZ TrA}, in only $42\,$minutes \citep{Middleditch+81,Chakrabarty98}. The optical companion is a very low-mass star (${<}0.1\,M_{\odot}$; \citealt{McClintock+77,McClintock+80}).
The Gaia measured distance to the star is $3.5^{+2.3}_{-1.3}\,$kpc, consistent with more recent measurements of $3.5^{+0.2}_{-0.3}\,$kpc by \citet{Schulz19}. Since its discovery, the source has shown two major torque reversal episodes.
The first was estimated to happen in 1990 \citep{Wilson+93, Bildsten+94} when the source switched from a steady spin-up trend, with a rate of $\dot{\nu}=8.5\times10^{-13}$\,Hz\,s$^{-1}$ that was observed for over a decade, to a $\sim7\,$yr long steady spin-down trend, with a rate of $\dot{\nu}=-3.5\times10^{-13}$\,Hz\,s$^{-1}$ \citep{Chakrabarty+97}.
The second torque reversal episode was observed by GBM\footnote{\url{https://gammaray.msfc.nasa.gov/gbm/science/pulsars/lightcurves/4u1626.html}} and Swift/BAT in 2008, when the source started a new spin-up trend. As of 2019 November, the source is still spinning up, with a mean rate of $\dot{\nu}=4\times10^{-13}$\,Hz\,s$^{-1}$ \citep{Camero+10}.

\subsubsection{Her X-1}
Her X-1 was discovered by Uhuru in 1971 \citep{Tananbaum1972}.
It is an eclipsing LMXB and one of the most studied accreting X-ray pulsars.
The NS spin period is about $1.2\,$s, while the orbital period is $\sim1.7\,$days.
The optical companion is an A7-type star \citep{Reynolds+97}, with a Gaia measured distance of $5.0^{+0.8}_{-0.6}\,$kpc, consistent with previous measurements of $6.1^{+0.9}_{-0.4}\,$kpc \citep{Leahy+14}.
The system exhibits super-orbital X-ray modulation with a period of $\sim35\,$days, most likely driven by a precessing warped accretion disk (see \citealt{Scott+00,Leahy+Igna10,Kotze+12} and references therein) or by a precessing NS \citep{Postnov+13}.
The general trend of the pulse period evolution is that of spin-up at an average rate of $\dot{\nu}=5\times10^{-13}\,$Hz\,s$^{-1}$ (see, e.g., \citealt{Klochkov09}). It occasionally exhibits spin-down episodes of a moderately larger entity, $\dot{\nu}=-7\times10^{-12}\,$Hz\,s$^{-1}$ \citep{Bildsten1997}.
Recently, GBM measurements\footnote{\url{https://gammaray.msfc.nasa.gov/gbm/science/pulsars/lightcurves/herx1.html}} have shown that the spin derivative trend has flattened around a spin frequency of $807.937\,$mHz.

\subsubsection{Cen X-3}

The bright, persistent source, Cen X-3, was discovered by the \emph{Uhuru} satellite in 1971 and marks the first observation of this accreting X-ray pulsar \citep{Giacconi1971}. The NS spin period is about $4.8\,$s, and the orbital period of the NS is $\sim2.1\,$days \citep{Kelley+83,Falanga+15} in an almost circular orbit ($e<0.0016$) around V779~Cen, an O6-7 supergiant companion star \citep{Krzeminski74,Hutchings+79, Ash99}.
The Gaia measured distance is $6.4^{+1.4}_{-1.1}\,$kpc, consistent with previous measurements of $5.7\pm1.5\,$kpc \citep{Thompson09}. Optical observations first revealed the presence of an accretion disk around the pulsar~\citep{Tjemkes+86}.
X-ray monitoring of the source later on found that the secular trend of the NS frequency is to spin-up; although, long spin-down periods have also been observed \citep{Nagase89CenX3}. GBM observations show that typical spin-up rates are of the order of $\dot{\nu}=3\times10^{-12}\,$Hz\,s$^{-1}$. Although the NS is accreting from a disk, there appears to be no correlation between the spin derivative and the observed X-ray flux in Cen~X-3 \citep{Tsunemi+96,Raichur+08a,Raichur+08b}. Instead, the high, aperiodic source variability is likely due to a radiatively warped accretion disk \citep{Iping+90} that does not reflect a real modulation of the accretion rate.
The most likely scenario is that the mass transfer in this system is dominated by RLO with wind-accretion (and wind-captured disk) contributions \citep[and references therein]{Walter15,ElMellah19}.
Orbital decay has also been observed for this source, at an average rate of $\dot{P}_{orb}/P_{orb}=-1.800(1)\times10^{-6}\,$yr$^{-1}$, and interpreted in terms of tidal interaction plus rapid mass transfer between the NS and its massive companion \citep{Nagase+92,Falanga+15}.
The long-term spin derivative is likely due to an accretion disk moving with alternating rotational direction, with a proposed periodicity in the Ginga data of about $9\,$yr \citep{Tsunemi+96}. However, the combined BATSE and GBM\footnote{\url{https://gammaray.msfc.nasa.gov/gbm/science/pulsars/lightcurves/cenx3.html}} spin frequency and pulsed flux history for this source reveal a more complex behavior and show alternating long-term spin-down and spin-up trends with random-walk variations superimposed \citep{deKool+93}.

\subsubsection{4U 1538--52}

Pulsations from 4U 1538--52 were first discovered in 1976 by the Ariel 5 mission, which revealed a spin period of $530\,$s \citep{Davison+77}.
The NS is in a $3.7\,$day, slightly eccentric ($e\sim0.2$) orbit \citep{Davison+77,Corbet+93,Clark00}.
The supergiant companion is a B0e-type star, called QV Nor \citep{Parkes+78}.
The Gaia measured distance is $6.6^{+2.2}_{-1.5}\,$kpc, consistent with a previous measurement of $5.8^{+2.0}_{-1.9}\,$kpc by \citet{Guver10}.
Early observations of this source show a long-term spin-down trend with random short-term variations \citep{Makishima+87,Nagase89}, while later observations find a reverse of the general long-term trend, probably happening in 1988. The average spin-up rate of $\dot{\nu}=1.8\times10^{-14}\,$Hz\,s$^{-1}$, is of the same order of magnitude as the previous spin-down period \citep{Rubin+97}. This spin-up period went on at least until 2006 \citep{Baykal+06}. As of 2019 November, GBM\footnote{\url{https://gammaray.msfc.nasa.gov/gbm/science/pulsars/lightcurves/4u1538.html}} has been observing the source in a new spin-down trend since the beginning of its operations in 2008. The binary source also shows hints of orbital decay, $\dot{P}_{orb}/P_{orb}=(0.4\pm1.8)\times10^{-6}\,$yr$^{-1}$ \citep{Baykal+06}, similar to the value observed in other X-ray binaries, yet consistent with a null value.

\subsubsection{Vela X-1}

X-ray pulsations from Vela X-1 were discovered by the SAS-3 satellite in 1975 \citep{McClintock+77}, revealing a spin period of about $283\,$s. The X-ray source is eclipsing, with an orbital period of $\sim8.9\,$days (\citealt{Falanga+15}, and references therein).
The orbit is almost circular ($e\sim0.09$) around HD77581, a B0.5Ia, supergiant \citep{Brucato+Kristian72,Hiltner+72,vanKerkwijk+95}. The Gaia measured distance of $2.42^{+0.19}_{-0.17}\,$kpc is consistent with previous measurements of $2.0\pm0.2\,$kpc \citep[and references therein]{Gimenez16}. The ellipsoidal variation of the optical light curve of the stellar companion suggests the star is distorted due to the tidal forces from the NS \citep[and references therein]{Koenigsberger12}.  
The NS is believed to be wind-fed as the evolution of the spin frequency does not show any steady long-term trend, but is instead observed to take a random-walk path \citep{Deeter+89,deKool+93}. However, the geometrical configuration of the binary system is such that a more complex phenomenology needs to be taken into account. Both simulated and observational studies (see, e.g., \citealt{Blondin+91,Kaper+94,Sidoli14,Malacaria+16}) have shown that three different structures are present in the binary: a photoionization wake due to the Str{\"o}mgren sphere around the NS, a tidal stream due to the almost filled Roche lobe of the donor, and a turbulent accretion wake in which transient accretion disks with alternating directions of rotation form around the NS (see also \citealt{Fryxell+Taam88,Blondin+12}). Recent works also show that wind-captured transient disks can form in the ambient wind of Vela X-1 (see, e.g., \citealt{ElMellah19}). Those transient accretion disks are thought to be responsible for the sparse spin-up/spin-down episodes observed by GBM at random epochs\footnote{\url{https://gammaray.msfc.nasa.gov/gbm/science/pulsars/lightcurves/velax1.html}}. 
The spin-up rate observed during such episodes is of the order of $\dot{\nu}=1-2\times10^{-13}\,$Hz\,s$^{-1}$, with  milder rates for spin-down episodes, $\dot{\nu}=-2\times10^{-14}\,$Hz\,s$^{-1}$.
With a magnetic field of $2.6\times10^{12}\,$G \citep[and references therein]{Fuerst+14} and an average luminosity of $5\times10^{36}\,$erg\,s$^{-1}$, and using $\Pi_{\rm su}=8$ from \citet{Shakura14a}, the quasi-spherical settling accretion model (Eq.~\ref{eq:QSAMup}) predicts a spin-up value of $\dot{\nu}=3\times10^{-14}\,$Hz\,s$^{-1}$, a factor of ${\sim}4$ weaker than what was actually observed.
On the other hand, the spin-up rate expected by accretion-disk theory (see Eq.~\ref{eq:parmar}) is $\sim1.4\times10^{-12}\,$Hz\,s$^{-1}$, almost an order of magnitude faster than the observed one.
Up to now, no clear correlation has been reported between the observed spin derivative episodes and the X-ray luminosity for this source.

\subsubsection{OAO 1657--415}

Pulsations from OAO 1657--415 were discovered by HEAO-1 with a spin period of about $38\,$s \citep{WhitePravdo1979}.
The orbital period is $10.4\,$days, with an eclipse occurring for $1.7\,$days \citep{Chakrabarty+93}. Accretion onto the NS is fed by an optical companion that is currently believed to be in a transitional stage between an OB and Wolf Rayet star, of the spectral type Ofpe/WN9 \citep{Mason2009}. The closest counterpart measured by Gaia is located at an angular offset of $4\arcsec.7$ from the nominal source position, at a distance of $2.2^{+0.7}_{-0.5}\,$kpc. 
However, previous measurements locate the system at a distance of about $4.4-12\,$kpc \citep{Chakrabarty02, Mason09}, consistent with the measurement of $7.1\pm1.3\,$kpc by \citet{Audley06}.
The pulsar exhibits a secular spin-up trend at an average rate of $\dot{\nu}\sim8.5\times10^{-13}\,$Hz\,s$^{-1}$, superimposed with several spin-up/spin-down episodes throughout its long history of observation. Analysis using BATSE and GBM\footnote{\url{https://gammaray.msfc.nasa.gov/gbm/science/pulsars/lightcurves/oao1657.html}} data allowed \citet{Jenke2012} to establish two different modes of accretion: one resulting from transient, disk-driven accretion that leads to steady spin-up periods correlated with flux (see also \citealt{Baykal97}), while the other results from wind-driven accretion, in which the NS spins-down at a (slower) rate that is uncorrelated with flux. Recently, a ``magnetic levitating disk'' scenario has been proposed to explain the spin evolution in OAO 1657--415 \citep{Kim+17}.

\subsubsection{GX 301-2}

GX 301-2 pulsations at $\sim680\,$s were discovered by Ariel 5 in 1976 \citep{White+76}.
The NS is in an eccentric ($e\sim0.5$), $41.5\,$day orbit \citep{Sato+86} around Wray~$77$, a hyper-giant, B1~Ia+ star \citep{Vidal73,Parkes+80,Kaper+95}. The measured Gaia distance is $3.1^{+0.6}_{-0.5}\,$kpc, consistent with a previously estimated distance of $3.1\,$kpc \citep{Hammerschlag+79,Kaper+06}. 
GX 301-2 is a near-equilibrium rotator; thus, its net spin derivative is equal to zero.
The secular spin period evolution is generally smooth and consistent with a random-walk evolution \citep{deKool+93}. 
However, the source has shown rapid ($\dot{\nu}=(3-5)\times10^{-12}\,$Hz\,s$^{-1}$) and prolonged ($\sim30\,$days) spin-up episodes \citep{Koh+97}, probably indicating the formation of a transient accretion disk.

Recently, GBM\footnote{\url{https://gammaray.msfc.nasa.gov/gbm/science/pulsars/lightcurves/gx301m2.html}} observed another similar episode over a longer period ($\sim40\,$days), which was also stronger ($\dot{\nu}\approx6\times10^{-12}\,$Hz\,s$^{-1}$) than was previously observed (\citealt{Nabizadeh19, Abarr20}; Malacaria C. et al. in preparation),
during which NuSTAR measured an unabsorbed luminosity of $\sim1.5\times10^{37}\,$erg\,s$^{-1}$ at a corresponding spin-up rate of $\dot{\nu}\approx3.6\times10^{-13}\,$Hz\,s$^{-1}$ as measured by GBM around the NuSTAR observation.
This is a factor of three lower than the spin-up rate predicted by Eq.~\ref{eq:QSAMup} of $\sim1.1\times10^{-12}\,$Hz\,s$^{-1}$.
On the other hand, the spin-up rate predicted by the accretion-disk theory (see Eq.~\ref{eq:parmar}), assuming a magnetic field of $4\times10^{12}\,$G \citep{Kreykenbohm04}, is $\sim4\times10^{-12}\,$Hz\,s$^{-1}$, about an order of magnitude stronger than the observed rate, but in agreement with the rate observed during the initial phase of the spin-up episode.

\subsubsection{GX 1+4}\label{subsubsec:gx1+4}

GX 1+4 is an LMXB discovered by a balloon X-ray observation in 1970 \citep{Lewin1971}, with an NS spin period of ${\sim}2\,$minutes \citep{Lewin1971}.
The orbital period is not yet well known. Old studies have reported periodic signals every $304\,$days in the optical band (e.g., \citealt{Cutler+86}) or at $1161\,$days in the X-ray band (e.g., \citealt{Hinkle+06}). However, RXTE data from GX~$1+4$ does not show modulation at any of these proposed periods \citep{Corbet+08}. The donor companion is \emph{V2116 Oph.} \citep{Glass1973}, a type M6III red-giant star that under-fills its Roche-Lobe \citep{Chakrabarty+Roche97,Hinkle+06}. Therefore, it is assumed that the NS is wind-fed by the companion, making it part of the so-called symbiotic X-ray Binaries (SyXBs; \citealt{Corbet+08, Yungelson19}). Different attempts have been made to determine the distance to the source (see, e.g., \citealt{Gonzalez+12} and references therein), but it still remains poorly constrained. The Gaia measured distance to the source is $7.5^{+4.3}_{-2.8}\,$kpc. Ten years following its discovery, the source was spinning up strongly at an average rate of $\dot{\nu}=6.0\times10^{-12}\,$Hz\,s$^{-1}$ \citep{Doty+81,Warwick+81, White+83}. No observations were recorded from 1980 and 1983, as EXOSAT did not detect the source, indicating that the flux had decreased below the sensitivity of the instrument. The source reappeared in 1987 in observations by Ginga with a lower luminosity and exhibiting a torque reversal with an average spin-down rate of $\dot{\nu}=-3.7\times10^{-12}\,$Hz\,s$^{-1}$
\citep{Makishima+88,Nagase89}.
With a magnetic field of $3.7\times10^{12}\,$G \citep{Ferrigno07}, and assuming an average luminosity of $4\times10^{36}\,$erg\,s$^{-1}$ (see \citealt[and references therein]{Gonzalez+12}), the quasi-spherical settling accretion model (Eq.~\ref{eq:QSAMdown}) predicts a comparable spin-down value, $\dot{\nu}=-1.5\times10^{-12}\,$Hz\,s$^{-1}$.
GBM\footnote{\url{https://gammaray.msfc.nasa.gov/gbm/science/pulsars/lightcurves/gx1p4.html}} observations show the source to still be undergoing a general spin-down trend, with occasional brief spin-up episodes corresponding to bright flux levels.

\section{Discussion}\label{sec:discussion}

The (almost) all-sky, continuous, long-term coverage of GBM provides fresh data that helps improve the analysis of accreting X-ray pulsars, providing enough statistics for interesting population studies. Examples of this are shown in Figures~\ref{fig:bimodality} and\ref{fig:torque}.

\subsection{Bimodal spin period distribution}

\begin{figure}[!t]
\includegraphics[width=0.45\textwidth]{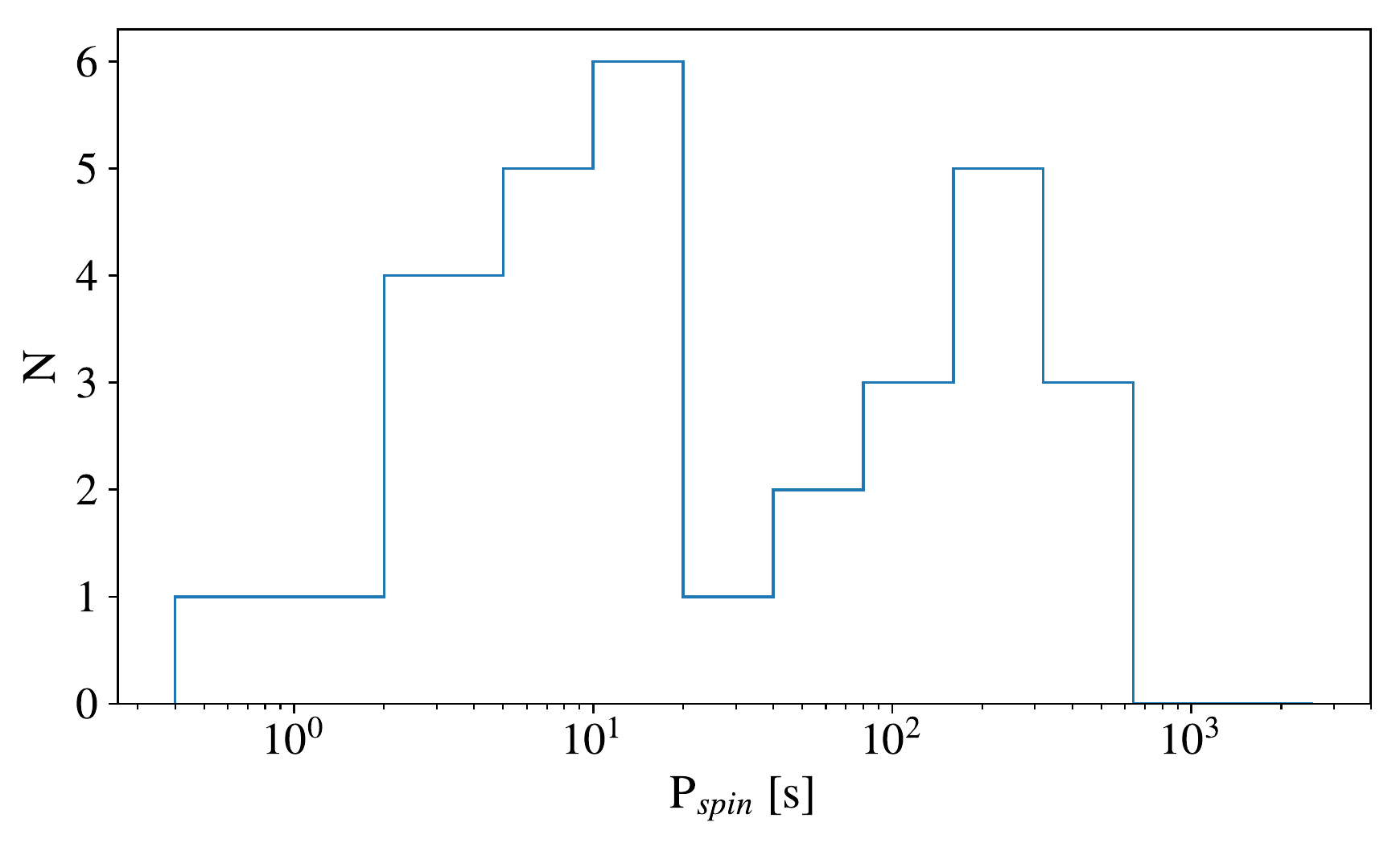}
\caption{The spin period distribution of all GBM detected BeXRBs in the Milky Way.}
\label{fig:bimodality}
\end{figure}

\citet{Knigge11} showed that the $P_{\rm s}-P_{\rm orb}$ distribution in BeXRBs is bimodal. 
While the bimodality of $P_{\rm orb}$ is only marginal, the distribution of $P_s$ has a clear gap at around $40\,$s and two distinct distributions peaking at $\sim10\,$s and $\sim200\,$s, respectively.
Those authors proposed that the two distinct subpopulations are created by the type of supernovae, that was the progenitor for the formation of the NS. One is an electron-capture supernovae (ECS), which would produce NSs with shorter spin periods on average, while iron core-collapse supernovae (CCS) would produce NSs with longer spin periods. Other explanations of the bimodality have also been described in previous studies. \citet{Cheng14} proposed that the bimodal distribution is the result of two different accretion modes: (1) advection-dominated accretion flow (ADAF) disks, which are more likely to form during a Type I outburst, or (2) thin accretion disks formed during a Type II outbursts. The ADAF disk is inefficient to spin-up the NS; thus, BeXRBs experiencing more Type I outbursts will produce NSs with longer spin periods. The thin disks produced as a result of Type- I outbursts are more efficient in transferring angular momentum, and thus, BeXRBs dominated by this mechanism will spin-up the NS to the shorter spin period subpopulation.

While those authors analyzed the cumulative population of BeXRBs (Milky Way, LMC, and SMC), it is of interest to test the bimodality for the Galaxy alone. GBM allowed us to constrain the ephemerides for a large number of such transients over the last decade, allowing us to probe the spin period distribution of BeXRBs in the Galaxy. This is shown in Fig.~\ref{fig:bimodality}, which shows two distinct distributions peaking at ${\sim}10\,$s and $\sim200\,$s with a clear gap at $\sim40\,$s, confirming the findings of \citet{Knigge11}.

\subsection{Accretion-driven torques}

\begin{figure*}[!t]\centering
\includegraphics[width=1.\textwidth]{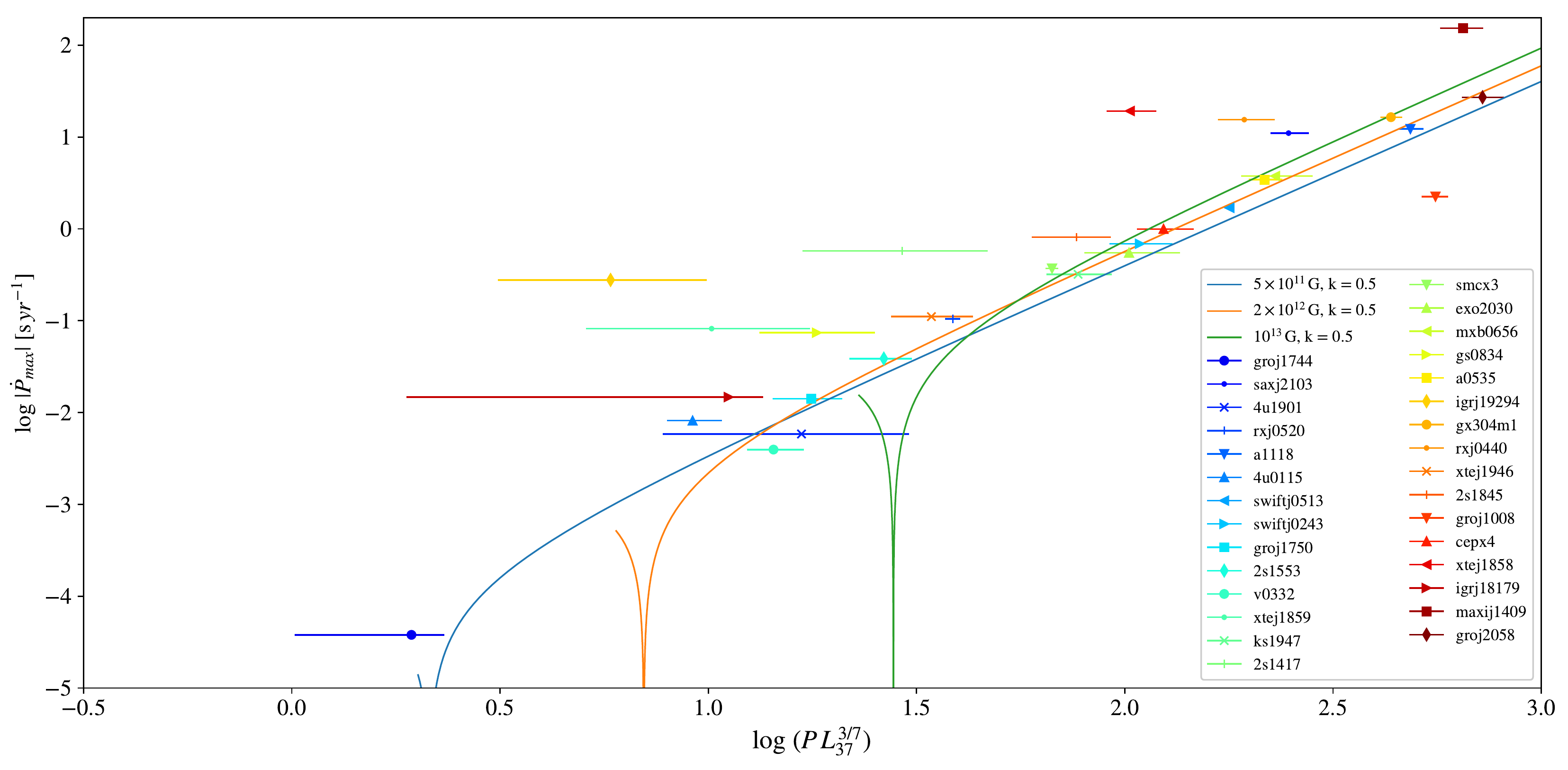}
\caption{Distribution of the maximum spin period derivative (absolute value) against the maximum peak bolometric luminosity normalized to the source spin period ($PL_{37}^{3/7}$) for all GBM transient sources.
The turquoise, orange, and green continuous lines correspond to the GL model (Eq.~\ref{eq:torque2}) for different magnetic field values, i.e. $5\times10^{11}$, $2\times10^{12}$ and $10^{13}\,$G, respectively (all assuming a coupling factor of $k=0.5$).
Model cuspids mark the value of the dimensionless parameter $n(\omega_s)=0$.
Despite the lack of an orbital solution, the following sources have been included in the plot: MXB~0656--072, GS~0834--430, IGR~J19294+1816, RX~J0440.9+4431, Cep~X-4, XTE~J1858+034, GRO~J1008--57, IGR~J18179--1621, MAXI~J1409--619, and GRO J2058+42.}
\label{fig:torque}
\end{figure*}

We analyzed the relationship between the spin-up strength and the observed luminosity for all of the GBM XRP transient sources. As a measure of the spin-up strength, we calculated the peak luminosity of the brightest outburst ever recorded over GBM lifetime for a given source and compared it with the corresponding maximum spin period derivative calculated over the same outburst. This allows us to study the dependence of these two parameters according to the Ghosh--Lamb (GL) model expressed by Equation~(\ref{eq:torque2}). For a meaningful comparison of the torque strength as a function of the luminosity, we considered the bolometric ($0.1-200\,$keV) luminosity inferred for each source as described in the Appendix~\ref{sec:bolo}.

Our results are shown in Fig.~\ref{fig:torque}, together with theoretical predictions from Eq.~\ref{eq:torque2}. The figure clearly shows a correlation, where the behavior of the measured sources mostly follows the predictions from the GL model. The model predicts different spin equilibrium values, $n(\omega_s)=0$, for different magnetic field strengths. The spin equilibrium is characterized by a cuspid in the GL function and discriminates between the fast rotator regime ($\omega_s>1$, to the left side of the cuspid) where the source is spinning down, and the slow rotator regime ($\omega_s<1$, to the right side of the cuspid) where the source is spinning up. Deviations of up to a factor of three have been highlighted by \citet{Sugizaki17} between observations of spin-up rates and predictions from the GL model.
However, the observed deviations were most likely ascribed to the uncertainty in the involved parameters, in particular the distance, $d$.
Thanks to the Gaia DR2 (see Appendix~\ref{sec:gaiadist}), we have been able to sensibly reduce the uncertainty on this parameter and compare the model predictions with the most precise data currently available. An example of how improved distances help with our understanding of source behavior,is especially pertinent to sources like V~0332+53 (X~0331+53). \citet{Sugizaki17} find that the deviation of this source from theoretical predictions is anomalously high, concluding that its behavior is due to an overestimate of the source distance. An improved distance obtained by Gaia ($5.1^{+1.1}_{-0.8}\,$kpc) was able to better constrain the source's behavior. Although marginal deviation is still present between observations and the torque model, the difference is significant only at the $1\sigma\,$c.l. While the the majority of the Gaia distances in this paper are consistent with previous measurements, there are some sources in this work that are inconsistent with distance values estimated through other methods, e.g., by adopting the luminosity dependence of the spin-up rate (see Equation~(\ref{eq:torque2})). Such a discrepancy is likely due to a combination of factors, including the uncertainty in the physical mechanisms driving the spin evolution of accreting NSs, the poorly constrained parameters of Equation~(\ref{eq:torque2}), and the limited range over which Gaia can measure distances. The latter implies that more reliable distances can be obtained for measured parallaxes that are smaller than their uncertainties, i.e. usually below $\sim5\,$kpc \citep{Bailer-Jones18, Luri18}.

Even considering the Gaia measured distances, a few sources still show considerable deviations from the GL model.
Some of these have unknown or poorly known orbital solutions, namely GS~0834--430, IGR~J19294+1816, RX~J0440.9+4431,  XTE~J1858+034, IGR~J18179--1621, MAXI~J1409--619, GRO~J1008--57 and GRO J2058+42. Part of the deviation is due to the lack of a timing solution; although, a few binary systems with known orbital solutions have also been observed to deviate from the GL model predictions, e.g., 4U~0115+634 and 2S 1417--624. Recently, \citet{Ji+2019} analyzed a 2018 outburst from 2S 1417-624 that employed a standard GL model to account for the spin-up shown by the source during the accretion episode.
However, they had to use a coupling parameter value of $k\sim0.3$ and a distance of $20\,$kpc in order to achieve a good fit. We find that a standard value of the coupling constant ($k\sim0.5$) and a measured Gaia distance of $\sim10\,$kpc is unable to account for the observed spin-up from that outburst (see Figure~\ref{fig:torque}), in agreement with findings from \citet{Ji+2019}.



\subsection{Improvement/Finding of orbital solutions}

Following the technique described in Sect.~\ref{sec:timing}, we derived or updated ephemerides for a selected sample of sources: 4U 0115+63, 4U 1901+03, and 2S 1553--542. To separate the pulsar emission from the background, good CTIME data are represented by Fourier components while a background model is fit to the data and then subtracted. The background model includes bright known sources, the variation in the detector responses, the Earth occultation steps, and a remaining long-term background contribution. Details on this procedure are described in \citet{Finger+99}, \citet{Camero+10}, and \citet{Jenke+12}.

The observed times are barycentered using the JPL Planetary ephemeris DE200 catalog \citep{Standish90}. The Barycentric Dynamical Time (TDB) are modulated by the binary orbital motion of the emitting pulsar, $t^{\rm em} = TDB - z$, and the binary orbital parameters can be constrained by fitting the TDB times to Equation~(\ref{eq:z_def}) (starting from an approximate, previously known solution). Orbital fits were obtained for two HMXBs due to their X-ray duty cycle. Best-fit orbital elements for those sources are presented in Table~\ref{tab:summary}.


\section{Summary}\label{sec:summary}
We have summarized more than $10\,$yr of Fermi/GBM accreting X-ray pulsars observations as part of the GBM Accreting Pulsars Program (GAPP). Detailed inspection of the spin history of accreting XRPs unveils a plethora of differences, highlighting the importance of continuous, wide-field monitoring observations. We showed how GBM observations are vital in addressing decade-long behaviors, such as long-term cycles in the Be disks (e.g., in EXO 2030+375) and torque reversals (e.g., in 2S 1845--024). Adherence of accretion-driven torques to the GL model for all GBM detected transient systems, as well as quasi-spherical accretion torques model predictions for a subsample of wind-accreting systems, have been tested, aided by updated source distances from Gaia DR2. This has allowed us to test the model predictions and to cross-check independent distance determinations (e.g., 2S 1417--624). Finally, we obtained new/updated orbital solutions for three accreting XRPs.
Our results demonstrate the capabilities of GBM as an excellent instrument for monitoring accreting X-ray pulsars and its important scientific contribution to this field.

\acknowledgments

We dedicate this paper to Dr. Mark Finger, retired, who initiated the GBM Accreting Pulsars Program and was an irreplaceable part of it. 
We thank the anonymous referee,  whose careful reading and  suggestions  have  improved  the  manuscript.
This research has made use of data and software provided by the High Energy Astrophysics Science Archive Research Center (HEASARC), which is a service of the Astrophysics Science Division at NASA/GSFC and the High Energy Astrophysics Division of the Smithsonian Astrophysical Observatory. We acknowledge extensive use of the NASA Abstract Database Service (ADS). C.M. is supported by an appointment to the NASA Postdoctoral Program at the Marshall Space Flight Center, administered by Universities Space Research Association under contract with NASA.

\facilities{Fermi/GBM, CGRO/BATSE, Swift/BAT} 
\software{HEASoft}

\bibliographystyle{yahapj}
\bibliography{references}

\appendix

\section{The bolometric luminosity}\label{sec:bolo}

The GL model provides the torque occurring on a source that accretes matter producing a certain luminosity. The luminosity considered in that model is a bolometric quantity in the X-ray domain, i.e. over the entire $0.1-200\,$keV energy band. However, it is rare that spectral data can offer such a wide energy coverage. Even so, the spectral energy distribution of accreting X-ray pulsars is rapidly folding beyond $\sim20\,$keV, and the flux measurement becomes progressively more uncertain at higher energies. Therefore, a different approach in deriving the luminosity at energies that differ from the band covered by the observing instrument must be taken.

A common approach is to evaluate the spectral model over a given energy range and then extend the it to cover the $0.1-200\,$keV energy range.
This approach assumes that the inferred spectral model shows no variations with respect to the model that was originally fit to the actual data. This is not always true, since additional spectral components sometimes emerge at different energies or luminosity states.
However, the spectrum between $1$ and $20\,$keV is thought to be the most representative because it is sufficient to characterize the cutoff power-law component, the column density absorption, and, possibly, an additional soft component (e.g., the blackbody component coming from the accretion disk).

In this work, we followed the same strategy. For each source, we first considered the best-fit models in the $1-20\,$keV found in literature (see Table~\ref{tab:summary}), describing the source spectrum at the highest observed luminosity. Successively, the model was extended to the bolometric energy band through the employment of a dummy response using \texttt{XSPEC}\footnote{The \texttt{dummyrsp} tool, \url{https://heasarc.gsfc.nasa.gov/xanadu/xspec/manual/node94.html}.}.
We are aware that this procedure introduces systematic uncertainty in the peak luminosity derivation, due to the possible emergence of additional spectral components at energies beyond the interval in which the best-fit model is calculated.
However, the estimated uncertainty in the flux derivation from this procedure is $15\%$ (see also \citealt{Zelati18}). This is a relatively low value when compared with the uncertainty in the luminosity calculation derived from the high uncertainty on the source distance that still holds for some sources.

\section{The Gaia distances}\label{sec:gaiadist}

On 2018 April 25, the Gaia Data Release 2 (DR2) was made available \citep{GaiaCollaboration18}.
Soon after, an online catalog\footnote{\url{http://gaia.ari.uni-heidelberg.de/tap.html}} was also made available by the Gaia team to list the distances for a large number of sources, based on the priors employed by \citet{Bailer-Jones18}.
To derive the most updated available distances for the sources in Table~\ref{tab:summary}, we adopted the following procedure:

\begin{itemize}
    \item[1.] Query the ``Single Source Search'' catalog for the source name as reported in Table~\ref{tab:summary} (This catalog is based on the SIMBAD astronomical database \citep{Wenger00}).
    \item[2.] Retrieve the corresponding Gaia ID. Only Gaia DR2 IDs have been used in this work (unless otherwise stated).
    \item[3.] Query the Table Access Protocol (TAP) for the distance of the correspondent Gaia ID, as recommended in Section~4 of \citet{Bailer-Jones18}. 
\end{itemize}

We report the angular offset between the closest Gaia counterpart and each source nominal position when the offset is larger than $3\arcsec.5$ (the typical systematic error for Swift/XRT positions\footnote{\url{https://www.swift.ac.uk/analysis/xrt/xrtcentroid.php}}).

\end{document}